%% file: main.tex
\documentclass[12pt]{article}
\usepackage[utf8]{inputenc}
\usepackage[letterpaper,margin=1in]{geometry}
\usepackage{libertine}
\usepackage[dvipsnames]{xcolor}
\usepackage{graphicx}
\usepackage[bf]{caption}
\usepackage{subcaption}
\usepackage[sort]{cite}
\usepackage[hyphens]{url}
\usepackage{siunitx}
\usepackage{authblk}
\usepackage[all]{nowidow}
\usepackage[italic]{mathastext}
\usepackage[hidelinks]{hyperref}

\title{General-Purpose Multicore Architectures}
\author{Saugata Ghose}
\affil{University of Illinois Urbana-Champaign \protect\\ Urbana, IL, United States}
\date{}

\newcommand{\todo}[1][]{}

\begin{document}

\maketitle

\begin{abstract}
The first years of the 2000s led to an inflection point in computer architectures:
while the number of available transistors on a chip continued to grow, crucial transistor scaling properties started to break down and result in increasing power consumption, while aggressive single-core performance optimizations were resulting in diminishing returns due to inherent limits in instruction-level parallelism.
This led to the rise of multicore CPU architectures, which are now commonplace in modern computers at all scales.
In this chapter, we discuss the evolution of multicore CPUs since their introduction.
Starting with a historic overview of multiprocessing, we explore the basic microarchitecture of a multicore CPU, key challenges resulting from shared memory resources, operating system modifications to optimize multicore CPU support, popular metrics for multicore evaluation, and recent trends in multicore CPU design.

\end{abstract}

\input{sections/intro.tex}

\input{sections/motivation.tex}

\input{sections/hardware.tex}

\input{sections/memory}

\input{sections/software}

\input{sections/metrics}

\input{sections/modern}

\input{sections/conclusion}

\section*{Acknowledgments}

Thanks to Ryan Wong, Sudhanshu Agarwal, Yiqiu Sun, and Minh S. Q. Truong for reviewing multiple versions of this chapter and providing helpful feedback.

{% \small
\newpage
\bibliographystyle{IEEEtranS}
\bibliography{refs}
}

\end{document}

%% file: sections/intro.tex
\section{Introduction}
\label{sec:intro}

\begin{figure}[t]
    \centering
    \includegraphics[width=0.75\linewidth]{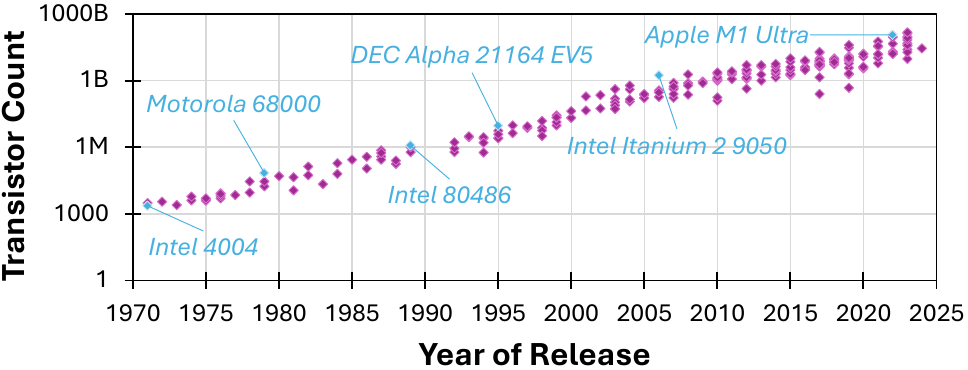}
    \caption{Log--linear plot of selected CPUs introduced between 1971 and 2024, illustrating the progression of Moore's Law.}
    \label{fig:moores}
\end{figure}

From the first commercial microprocessors in the 1970s through the end of the 1990s, microprocessors available on the market typically consisted of a single CPU core per chip. 
During that time, significant architectural advancements were made to improve the performance of the CPU core, including (but not limited to) techniques such as out-of-order processing and superscalar execution that extracted \emph{instruction-level parallelism} (ILP) from single-thread sequential programs.
These architectural advancements were driven by two trends that governed advances in semiconductor manufacturing process technologies.
The first, Moore's Law\cite{moore.electronics1965}, was an observation made in 1965 by Gordon Moore that the number of transistors on an integrated circuit (IC) doubled every year, which he revised in 1975~\cite{moore.iedm1975} to forecast a doubling every two years after 1980.
Figure~\ref{fig:moores} shows a progression of Moore's Law, using real CPUs as examples, between 1971 and 2024.
The second, Dennard scaling~\cite{dennard.iedm1972, dennard.jssc1974}, was a relationship identified by Robert Dennard and his colleagues at IBM in the early 1970s, identifying that with every new manufacturing process technology node (an approximately 18-month interval at the time), both the area and the power consumption of a single transistor was half of what was observed in the previous generation.

As Moore's Law and Dennard Scaling made it more economical to increase the number of transistors on a chip (effectively providing double the number of transistors, at the same area and power budget, every 18--24 months), manufacturers dedicated these additional transistors towards increasing the performance of the single CPU core.
Unfortunately, two critical factors made it difficult to keep continuing this trend.
First, as the ``free'' rewards of scaling started to break down in the early 2000s, the areal power density (directly correlated to the amount of heat dissipated per unit area) of high-end CPUs began growing rapidly~\cite{de.dac1999, frank.procieee2001}. Then-contemporary projections estimated that if single-core CPU development continued along the trends of the time, commonplace passive cooling elements would no longer be able to dissipate the heat generated by the CPU~\cite{de.dac1999}. 
Second, more aggressive techniques for ILP were yielding diminishing returns, requiring high hardware costs for meager performance benefits~\cite{ronen.procieee2001}.
These factors forced manufacturers to reconsider how to use additional transistors to continue to deliver performance improvements, now in a power-efficient manner~\cite{de.dac1999, ronen.procieee2001, esmaeilzadeh.isca2011}.

This reconsideration led to the widespread adoption of \emph{multicore CPUs} (also known as \emph{chip multiprocessors})~\cite{ronen.procieee2001, esmaeilzadeh.isca2011}: instead of trying to make a single CPU core more powerful, manufacturers now implement multiple, simpler CPU cores within a single chip that can run multiple tasks concurrently.\footnote{This chapter uses the term \emph{concurrent processing} to refer strictly to multiple independent tasks executing at the same time. It refers to the time multiplexing of a CPU core across multiple tasks as \emph{time-sharing}.}
The simpler core designs significantly reduced the areal power density (e.g., \si{\watt \per \milli\meter\squared}) of the CPUs.
Multicore CPUs perform concurrent processing by 
(1)~performing concurrent \emph{multiprocessing} of more than one program, and/or 
(2)~extracting \emph{thread-level parallelism} (TLP) from a parallelizable application.
While initial commercial multicore CPUs started out with two identical CPU cores, today's multicore CPUs have a wide range of configurations, with some containing dozens of cores.

This chapter will examine six topics related to multicore CPU architectures.
First, it will motivate the benefits and limitations of processing multiple tasks concurrently, and how they drove the need for parallel processing.
Second, 
it will examine the hardware design of a typical multicore CPU, and the changes required to support the efficient execution of multiple programs on multiple CPU cores.
Third, it will study how memory management policies change to handle the increased traffic from multiple cores.
Fourth, it will address software issues that optimize the use of multicore CPUs.
Fifth, 
it will introduce common metrics that are used in the context of multicore CPUs.
Finally, it will close by briefly discussing how commercial multicore CPUs have evolved since their introduction.

%% file: sections/motivation.tex
\section{Motivating the Need for Concurrent Processing}
\label{sec:motivation}
\label{sec:bkgd}

As early as the days of analog computing, the benefits of parallelizing tasks became a clear goal:
Luigi Federico Menabrea, in his 1842 study of Charles Babbage's Analytical Engine~\cite{menabrea.bibliothequeuniverselle1842}, noted that a key advantage of mechanized computation would be its ability to produce several results at the same time.
In the mid 20th century, as digital computers evolved rapidly, there was a pressing need to maximize both the performance and utilization of these computers, and the extraction of parallelism, in its various forms, became a key approach to meet these needs.
While this chapter focuses on multicore CPUs, a brief taxonomy of parallel computing hardware is provided in Section~\ref{sec:motivation:taxonomy}.

The early decades of electronic computing saw the introduction of several now-commonplace parallelization techniques, such as instruction pipelining (1941, with the Zuse Z3~\cite{rojas.annalshistory1996, zuse.patent1949}, followed by significant advances in 1961 with the IBM 7030 Stretch~\cite{dunwell.ejcc1956} and ILLIAC II~\cite{taub.tr1957} supercomputers), superscalar processing (1964, with the CDC 6600 supercomputer~\cite{thornton.fjcc1964}), and general-purpose time-sharing (1961, with the Compatible Time-Sharing System operating system~\cite{corbato.sjcc1962}).\footnote{In this section, significant efforts were made to identify milestone systems that introduced or made critical leaps forward in now-fundamental parallelism techniques. However, given the rapid pace of concurrent computer development in the 1950s and 1960s, combined with limited available historical resources, these examples may not always be the true originators of the techniques. When possible, dates associated with systems denote the year of first delivery, as a proxy for the completion of the first working system.}
This set the stage for the rise of the two key types of parallelism exposed by multicore CPUs: multiprocessing (Section~\ref{sec:motivation:multiprocessing}) and thread-level parallelism (Section~\ref{sec:motivation:tlp}).
While both of these types of parallelism can be exploited by a range of hardware organizations, historical trends in chip manufacturing process technologies drove the emergence of the multicore CPU (Section~\ref{sec:motivation:history}).

\subsection{Classifying Parallel Computing Hardware}
\label{sec:motivation:taxonomy}

\begin{figure}[h]
    \centering
    \includegraphics[width=0.25\linewidth]{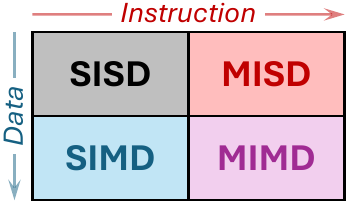}
    \caption{An illustration of Flynn's taxonomy.}
    \label{fig:flynn}
\end{figure}

Given the rise of multiple parallel computing techniques throughout the 1960s, there was a need to categorize the techniques based on broadly-shared principles.
To this end, Michael Flynn developed what is now known as Flynn's taxonomy in 1966~\cite{flynn.procieee1966}.
The original taxonomy classified computer architectures into four categories, along two dimensions as shown in Figure~\ref{fig:flynn}:
\begin{itemize}
    \item Single instruction stream, single data stream (SISD)
    \item Single instruction stream, multiple data stream (SIMD)
    \item Multiple instruction stream, single data stream (MISD)
    \item Multiple instruction stream, multiple data stream (MIMD)
\end{itemize}

Under Flynn's taxonomy, non-pipelined, non-superscalar single-core CPUs that employ a von Neumann architecture are a canonical example of a SISD architecture.
Examples of SIMD architectures include graphics processing units (GPUs), array processors, and vector extensions for CPUs.
One reasonably-accepted example of a MISD architecture is redundant computing, such as the systems employed in avionics units.\footnote{There is some latent controversy about which category pipelined CPUs fall into. In its best imitation of Switzerland and its neutrality, this chapter does not take a stand for any category. Sorry for being so bland.}
The MIMD category includes most computers that are capable of multiprocessing (Section~\ref{sec:motivation:multiprocessing}), including computers with multicore CPUs.

In a multicore CPU, each CPU core is capable of executing one or more independent streams of instructions, with each instruction stream operating on its own data set (though this data set can potentially overlap with the data sets being used by other cores).
To support this MIMD model of execution, each core has its own control logic, and much of the computation takes place independent from other cores.
Section~\ref{sec:hw} elaborates on this more, while also discussing when coordination between cores does occur (both explicitly and implicitly).

\subsection{Multiprocessing}
\label{sec:motivation:multiprocessing}

Before the days of personal computers (PCs), mainframes were the dominant form factor of computers. Mainframe computers had a relatively high cost,\footnote{The IBM System/360 Model 25, a low-end variant of IBM's highly-popular mainframes, was announced in January 1968 with a purchase price of US \$253,000 (US \$2.33 million in 2024 dollars), with a monthly rental option at \$5,330 (US \$49,000 in 2024 dollars)~\cite{whitney.moderndata1968}.} and given the high demand to use these computers, batching and time-sharing (i.e., time multiplexing) systems became commonplace. Systems capable of batching would queue up multiple jobs to execute on a computer back-to-back, while time-sharing systems extended this capability to execute multiple programs on the same computer through temporal multiplexing.

More concretely, time-sharing involves the use of \emph{scheduling quanta}, which are (typically predetermined) periods of time that a program can execute for before being \emph{preempted} (i.e., switched out). A time-sharing system runs one program for a single scheduling quantum, after which it switches out the program from the CPU and looks at a queue of pending programs (which include the one just switched out, assuming it did not complete within the quantum). The system then selects the next program to run, switches it into the CPU, and lets it execute for a single scheduling quantum, before repeating the switch-out/switch-in procedure. Time-sharing allowed multiple users to interact with a single computer at the same time, allowing each user's program to make forward progress even though the computer had only a single CPU. Thanks to the typically short scheduling quanta (on the order of milliseconds in modern machines), time-sharing often gives each user the illusion that the computer is continuously executing their application, as the scheduling quantum is significantly faster than human perception of response time.

While time-sharing represents a key technological shift in the accessibility of computers, the technique suffers from two critical overheads that significantly extend the overall execution time of a program.
First, in an ideal machine, each program now takes longer to complete as it repeatedly waits on other programs to execute. For example, if a computer runs 10~programs, each taking the same amount of time to finish, and the operating system uses round-robin scheduling to switch between each program, the programs will take 10$\times$ longer to complete than if they ran uninterrupted.
Second, in real-world machines, the execution time of each program is further increased by the overhead of \emph{context switching}. When a program is switched out, its registers (and for some machines, dirty cache values) must be saved somewhere, and these values must be restored when the program is switched back in.

As a natural progression towards concurrent execution without these overheads, researchers began to explore whether a computer could incorporate multiple CPUs, where the CPUs have the ability to communicate with each other.\footnote{Alternative techniques such as multithreading were also being developed around this time, with 1960's Bull Gamma 60~\cite{dreyfus.wjcc1958} and Honeywell 800~\cite{honeywell.800.manual.1960} being two early examples of hardware multithreading support. A constrained version of multithreading was implemented in the DYSEAC in 1954~\cite{leiner.tr1952}.} This technique, known as \emph{multiprocessing}, was first implemented in the Burroughs B~5000~\cite{longergan.datamation1961} and D825~\cite{anderson.fjcc1962} mainframe computers, released in 1961 and 1962, respectively. 
In a computer capable of multiprocessing, multiple programs can execute simultaneously, without the need to time-share the CPU. An operating system (OS) typically sees a multi-CPU computer as a single system, and a multiprocessing-capable OS is responsible for assigning programs to specific CPUs. Given the high cost of CPUs, many multi-CPU machines still perform time-sharing on each CPU in addition to multiprocessing, and the OS uses various scheduling heuristics to determine how best to choose which CPU a program should be scheduled on.

\subsection{Thread-Level Parallelism (TLP) Within an Application}
\label{sec:motivation:tlp}

While multiprocessing enabled increases in the throughput and/or productivity of a machine by allowing many applications to execute concurrently, there still remained a need to reduce the execution time of a single application.
Hardware techniques such as instruction pipelining and superscalar execution, along with compiler assisted techniques such as very long instruction words (VLIW), unlocked \emph{instruction-level parallelism} (ILP) within an application, by identifying independent instructions that can execute concurrently. 
While ILP techniques have provided significant performance boosts within a single process, there was also a need to exploit multiprocessing-capable machines to run a \emph{parallel program}, where multiple constituent tasks from the program execute concurrently.

In order for a parallel program to be decomposed into multiple parallelizable tasks, there needs to be
(1)~some way for these tasks to execute concurrently for much of their lifetime, as well as
(2)~some way for the tasks to communicate and synchronize with each other in order to generate a single set of results.
Parallel programs require support at both a software and hardware level to execute.
In software, the core concepts on how to decompose a program into multiple parallel tasks were developed in the 1950s and 1960s, and Dekker's algorithm from 1960~\cite{dijkstra.ewd35.note1962, dijkstra.bookchapter1968, dekker.bookchapter2022} allowed programmers to enable mutually exclusive access to shared memory locations among two concurrent processes. 
In hardware, the development of early multi-CPU machines was accompanied with the introduction of shared memory. For example, the Burroughs B~5000 from 1962~\cite{longergan.datamation1961} provided primitive shared memory support by letting all of its CPUs have shared access to its memory modules. 
As the concept of parallel programming evolved, these tasks eventually became formalized as \emph{threads} that can execute asynchronously~\cite{witt.ibmsys1966}, and the use of multiple concurrent threads to accelerate the execution of a program became known as \emph{thread-level parallelism} (TLP).

\begin{figure}[h]
    \centering
    \includegraphics[width=0.65\linewidth]{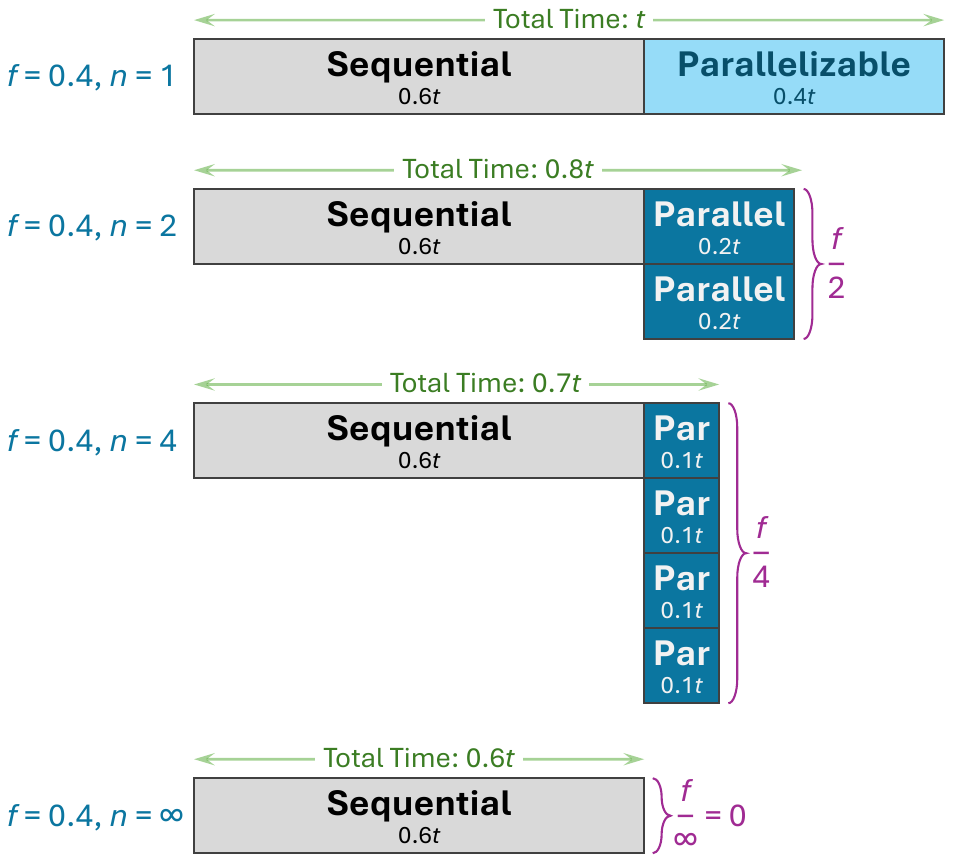}
    \caption{Amdahl's Law visualized for an example application.}
    \label{fig:amdahlexample}
\end{figure}

The performance improvement that a parallel program can achieve over its sequential counterpart depends primarily on two factors:
(1)~the number of threads that can be executed concurrently, and
(2)~the amount of synchronization needed.
Amdahl's Law models the impact of the thread count on maximum performance improvements for a \emph{fixed} problem size~\cite{amdahl.sjcc1967}.
The law states that the theoretical \emph{speedup} $S$ (i.e., the reduction in execution time) of a parallel program over its sequential counterpart can be expressed as:
\begin{equation}
    S = \frac{1}{(1-f) + \frac{f}{n}}
\end{equation}
where $f$ is the fraction of the program that can be parallelized, and $n$ is the number of concurrent threads that can execute the parallelizable part of the program.
To visualize Amdahl's Law, Figure~\ref{fig:amdahlexample} shows an example where $f = 0.4$ (i.e., 40\% of the program can be parallelized), for various thread counts.
As the figure shows, for every doubling of the thread count, the reduction in total time is only half of the reduction achieved by the previous doubling.

\begin{figure}[ht]
    \centering%
    \begin{subfigure}[b]{0.49\linewidth}
        \includegraphics[width=\linewidth]{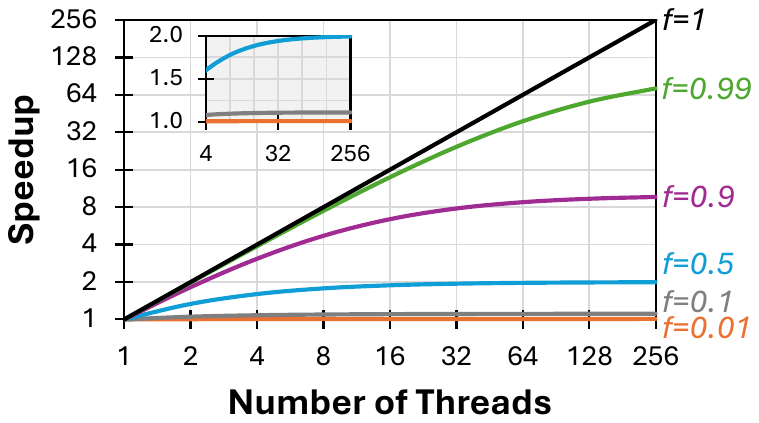}
        \subcaption{Amdahl's Law.}
        \label{fig:laws:amdahl}
    \end{subfigure}%
    \hfill%
    \begin{subfigure}[b]{0.49\linewidth}
        \includegraphics[width=\linewidth]{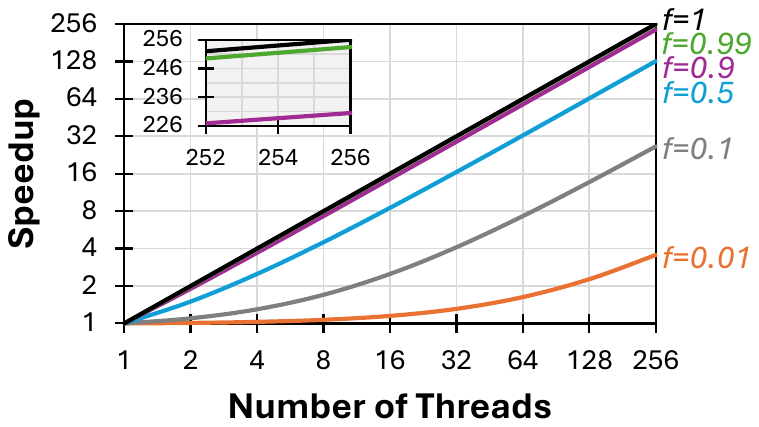}
        \subcaption{Gustafson's Law.}
        \label{fig:laws:gustafson}
    \end{subfigure}%
    \caption{Comparison of theoretical parallel speedup estimated by Amdahl's Law and Gustafson's Law, for different parallelizable fractions $f$. Inset graphs show a zoomed-in section of the main graph for clarity.}%
    \label{fig:laws}%
\end{figure}

Figure~\ref{fig:laws:amdahl} shows the theoretical parallel speedup achievable according to the law, for different values of $f$. 
A key takeaway from Amdahl's Law is that even for an infinite number of concurrent threads (i.e., $n=\infty$), the maximum speedup $S$ of a parallel program is bound by the time it takes to execute the part of the program that is \emph{not} parallelizable.
While Amdahl's Law encompasses key properties of parallel execution, it has two important limitations.

First, Amdahl's Law does not explicitly account for the additional overhead of synchronization primitives. As more threads execute concurrently, the contention for acquiring mutually exclusive access to a portion of shared memory can increase.
For example, if all of the threads of a parallel program share and update a single counter, they must use a mutual exclusion primitive (i.e., \emph{mutex}) to ensure that updates from one thread are not inadvertently lost by another thread.
This requires each thread to acquire the mutex whenever it updates the counter, and other threads that attempt to acquire a mutex that is currently held by another thread will stall. As more threads execute concurrently, the likelihood increases that more threads will contend at the same time to acquire the mutex, introducing more stalls. In practice, these synchronization overheads from mutex contention (as well as interference between threads due to resource sharing; see Section~\ref{sec:hw:mem}) can make it such that adding a parallel thread can actually \emph{decrease} the parallel speedup, as shown in Figure~\ref{fig:realspeedup} below.

Second, Amdahl's Law assumes that the problem size is fixed regardless of the number of parallel threads.
In reality, the number of threads used to parallelize a program is typically linked with the number of available CPUs.
For a multiprocessing machine, when there are more CPUs, a program has more resources available to it, such as memory capacity.
As a result, programmers and/or users can increase the problem size to take advantage of these extra resources.
Since Amdahl's Law does not capture the impact of these extra resources, it can provide a pessimistic estimate of the capabilities of a multi-CPU machine.
Gustafson's Law~\cite{gustafson.cacm1988} accounts for this, by calculating the parallel speedup $S$ as:
\begin{equation}
    S = (1 - f) + f \times n
\end{equation}
Figure~\ref{fig:laws:gustafson} shows the predicted speedups from the law, to provide a comparison to Amdahl's Law.

Real-world parallel system performance deviates from both Amdahl's Law and Gustafson's Law. For example, while one could treat Gustafson's Law as an optimistic upper bound on performance, real machines can sometimes outperform the estimated parallel speedup from Gustafson's Law due to data sharing. If one thread brings data into a shared memory that is subsequent used by a second thread (exploiting temporal and/or spatial locality across threads), the second thread no longer pays the memory latency required for that subsequent data access (assuming that the data is not evicted). This phenomenon is an example of \emph{super-linear speedup}.

Putting this all together, Figure~\ref{fig:realspeedup} shows a synthetically-constructed example of the parallel speedup one could expect to observe on a real parallel system.
The example assumes that 99\% of the application can be parallelized across all threads, while the remaining 1\% must execute sequentially.
It also assumes that the machine has 32 CPU cores, and since it has a fixed set of available resources, Amdahl's Law can be used to predict performance.
There are three observations from the figure.
First, even with a single thread, the observed speedup is less than 1.
This is because parallel speedup is compared against the best sequential implementation (see Section~\ref{sec:metrics}), and a parallel implementation (with a configurable parameter $n$ that sets the thread count) typically has overheads associated with adding code to support parallel execution, and these overheads are observed when $n = 1$.
Second, super-linear speedups can even exceed perfect parallelism.
As mentioned above, this is the result of threads helping each other through data locality.
Third, after a certain thread count (24 in our example), the performance at a higher thread count is actually \emph{lower} than the performance at a lower thread count.
This is because the benefits of additional thread-level parallelism and additional data locality are overcome by synchronization and interference overheads.

\begin{figure}[h]
    \centering
    \includegraphics[width=0.45\linewidth]{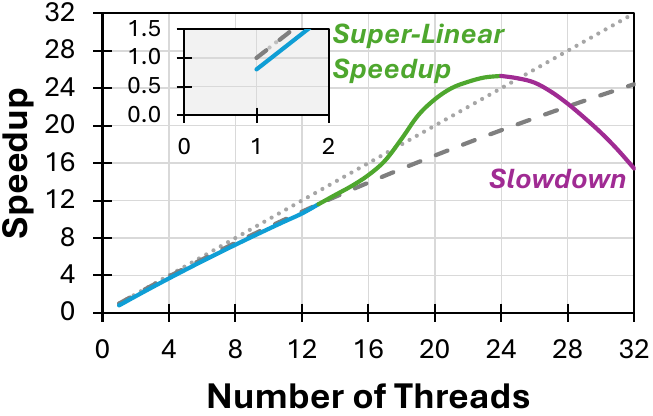}
    \caption{A synthetic example of observable behavior in a real-world parallel system. The dotted line shows ideal parallelism ($f=1$ with Amdahl's Law), the dashed line shows expected performance ($f=0.99$ with Amdahl's Law), and the solid line shows observed performance.}
    \label{fig:realspeedup}
\end{figure}

Note that while Figure~\ref{fig:realspeedup} is one example of parallel application behavior, its trends are not universal.
As one example, some applications exhibit what is known as \emph{embarrassingly parallel} behavior~\cite{mattson.book2004}, where an application continues to achieve near-ideal parallel speedup even at high thread counts, due to limited need for synchronization and serialization.

\subsection{What to Do With All These Transistors?}
\label{sec:motivation:history}
\label{sec:history}

Through the 1990s (and drawing some very broad, and likely inaccurate, generalizations), the vast majority of parallel computing architectures catered to the large-scale computing market, with much of the effort focused on supercomputers.
Most personal computers (PCs) and workstations\footnote{At that time, there was a distinction between more mainstream PCs and high-performance workstations that targeted power users. Today, in large part due to advancements in PC capabilities throughout the 1990s, this distinction has mostly disappeared, but the chapter references it here for historical context.} incorporated only a single-core CPU (or in the case of some high-end workstations, multiple single-core CPUs installed on a multi-socket motherboard).
Much of the focus on architectural innovation remained on improving the performance of the single core, leading to the maturation and widespread commercial adoption of techniques such as out-of-order processing and superscalar execution.

Concurrently, several architects started envisioning a new direction of CPU design.
In the early 1990s, with no immediate end in sight to Moore's Law and Dennard Scaling, a number of works speculated that with millions of additional transistors becoming available on a chip within the next decade, it would now be feasible to \emph{replicate multiple CPU cores within a single chip}.
This concept came to be known as a
\emph{single-chip multiprocessor} (which ultimately led to the term \emph{chip multiprocessor}, or CMP).
A number of varying designs were proposed, 
ranging from asymmetric and/or specialized cores~\cite{joyce.patent1987, schmidt.ieeemicro1991} to clustered CPUs~\cite{sohi.isca1995, fillo.micro1995}, as well as an early symmetric core design with a shared cache~\cite{hanawa.iccd1991}.
The Hydra CMP project at Stanford started to explore the trade-offs of using \emph{shared caches} across multiple smaller CPU cores for coordination, in comparison to a high-performance single-core CPU, which led to their seminal 1996 paper~\cite{olukotun.asplos1996} that advocated for many of the key features of a modern multicore CPU. 
Around that same time, architects at IBM started designing what would become the first commercially-available multicore CPU for non-embedded computers, the POWER4~\cite{tendler.ibmjrd2002}.\footnote{The earliest known implementation of a commercial multicore single-chip CPU is the COP2440 series by National Semiconductor~\cite{national.cop2440.datasheet.1982} (along with the COP2404 variant for prototyping), which was released in 1982. The COP2440 combined two embedded CPU cores on a single die to enable concurrent processing of real-time operations, with a shared memory subsystem.}
The POWER4 combined two CPU cores onto a single die,\footnote{Unfortunately, the term chip has overloaded definitions, so this chapter will use the following. A \emph{die} is a piece of semiconductor material (typically silicon), and a \emph{chip} is an integrated circuit that consists of one or more dies. A chip is typically placed inside a \emph{package}, which allows the chip to be placed in a socket to interact with other chips and electrical components.}
with a core frequency of \SI{1.1}{\giga\hertz} and a \emph{thermal design point} (TDP) of \SI{115}{\watt}.

The wider adoption of multicore CPUs was hastened by two trends observed in the late 1990s.
First, a handful of computer engineers started raising the alarm on an impending crisis with power consumption.
While Moore's Law (focused on increasing transistor count) remained alive and well (in what was truly a simpler time for the architecturally inclined), they predicted that Dennard scaling would begin to break down in the near future, which when combined with aggressive architectural changes would significantly increase both the power consumption of a chip and the chip's areal power density~\cite{de.dac1999, frank.procieee2001}.
As one example, in the span of a decade, Intel saw the the TDP\footnote{Conventionally, the thermal design point of a CPU represented its maximum power consumption, and served as a direct proxy for the maximum amount of heat that the CPU would dissipate. In today's usage, however, TDP may not account for additional power consumed when turbo modes are engaged, and TDP instead represents the longer-term maximum power under non-turbo steady state.} of its high-end CPUs increase by an order of magnitude, from \SI{3.5}{\watt} for the i486DX-25~\cite{intel.i486.manual.1989} (released in 1989) to \SI{34.5}{\watt} for the Pentium III 600~\cite{intel.piii600.ark.1999} (released in 1999).
Then-contemporary projections expected the thermal density of a chip to double every four years, with the density approaching that of a nuclear reactor by the mid-2000s~\cite{de.dac1999, ronen.procieee2001}.
While the industry still had increasing transistor counts for a fixed area of silicon, it was becoming more difficult to use all of these transistors to their full capacity without surpassing the capabilities of thermal dissipation techniques.
Second, more aggressive techniques for extracting instruction-level parallelism (ILP), such as deeper pipelining, larger superscalar widths, and deeper instruction lookahead for out-of-order processing, were resulting in diminished returns~\cite{ronen.procieee2001}.
These limits arose due to a combination of
(1)~inherent limits to the amount of ILP available in modern applications (due to dependencies and data-dependent control flow) and 
(2)~the quickly increasing complexity of logic required to extract further ILP. \todo[expand]
Combined, these issues made it increasingly difficult to continue aggressive single-core performance scaling, and multicore CPUs presented an attractive alternative to improving system performance.

As the breakdown of Dennard scaling materialized during the first few years of the 2000s, the power consumption issue was exacerbated significantly.
This added pressure led the next wave of multicore CPUs to make a key trade-off to avoid the impending limits of thermal density: they operated each CPU core at a \emph{lower} frequency than their single-core contemporaries, in the hopes that the increased parallel processing capability would make up for reduced single-core performance~\cite{esmaeilzadeh.isca2011}.
Section~\ref{sec:hw:cores} discusses more about the CPU core trade-offs, which allowed architects to harness TLP instead of aggressive ILP to improve performance while keeping power issues under control, and opened the floodgates for the proliferation of multicore CPUs.

%% file: sections/hardware.tex
\section{Multicore CPU Hardware Design}
\label{sec:hw}

While the term \emph{multicore CPU} may lead one to think that there are significant additions at a microarchitectural level to enable parallel execution, the central design of the cores are often (but not always) \emph{simpler} than the most aggressive single-core CPU microarchitectures released by manufacturers. 
In fact, the majority of the critical hardware changes to enable multicore CPUs lie \emph{outside} of the CPU core.
A central tenet of multicore CPU design is that performance can be attained through parallelism and repetition, as opposed to design complexity.

\begin{figure}[h]
    \centering
    \includegraphics[width=0.55\linewidth]{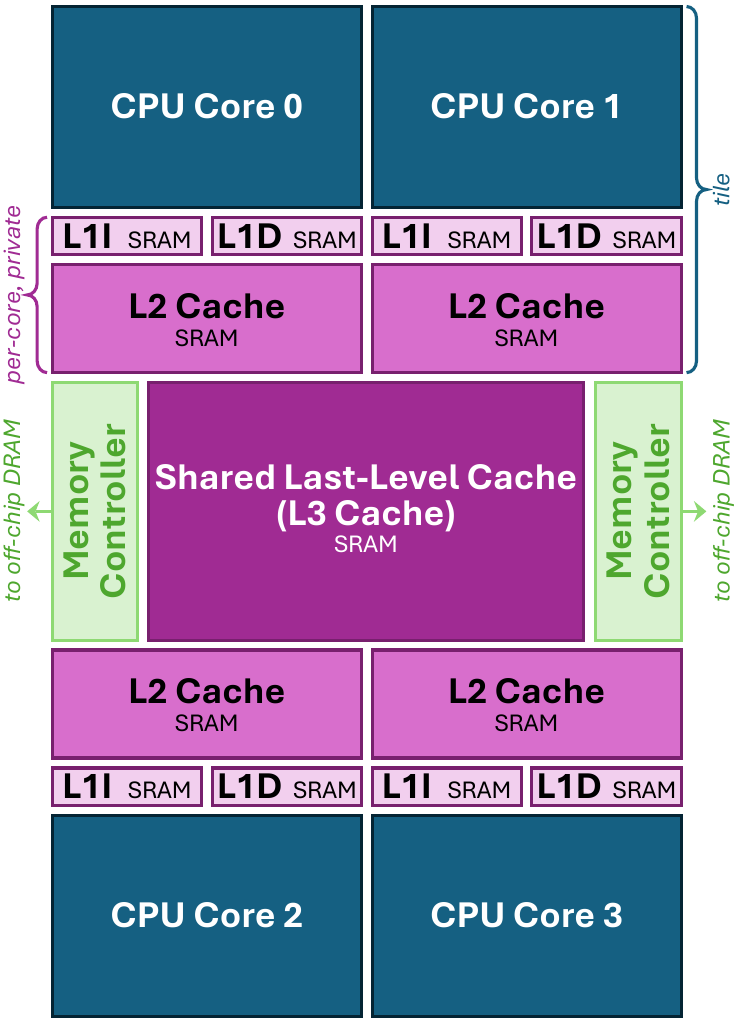}
    \caption{An example four-core multicore CPU.}
    \label{fig:multicore}
\end{figure}

Figure~\ref{fig:multicore} shows a typical example of the major components found in a multicore CPU.
Two or more CPU cores sit within the same chip package, with each core having its own (i.e., \emph{private}) L1 instruction, L1 data, and unified L2 caches.
All of the cores have \emph{shared} access to a single \emph{last-level cache} (LLC).
The LLC connects with one or more on-chip memory controllers, which provide access to off-chip DRAM modules in the system.

\subsection{Optimizing CPU Cores for Parallelism}
\label{sec:hw:cores}

Multicore CPU designers prevent the design and verification of the CPU from increasing linearly with the core count by relying on design reuse through \emph{tiling}.
In a homogeneous multicore CPU, the design of the cores is identical, in order to reduce manufacturing and verification/testing costs.
(Section~\ref{sec:modern} discusses popular approaches for heterogeneous multicore CPU design.)
Figure~\ref{fig:multicore} shows how the majority of the chip for a typical homogeneous multicore CPU is made up of multiple tiles, where each tile has exactly the same components.
A tile usually consists of a single CPU core and the private caches associated with that core.
Some architectures include a slice of the last-level cache in the tile (see Section~\ref{sec:hw:llc}).
In addition to these tiles, the chip includes shared components (e.g., controllers that connect the CPU cores to external components, interconnects).

To enable multiple tiles to be stamped out without incurring an exorbitant increase in chip area compared to a powerful single-core CPU, the CPU core within the tile is significantly simplified.
As an example, the Core 2 Duo, the second line of desktop multicore CPUs by Intel (and the first whose microarchitecture was designed with dual-core processing in mind), was not based off of the then-state-of-the-art NetBurst microarchitecture~\cite{boggs.itj2004} that formed the basis of the Pentium~4. Instead, it utilized a new microarchitecture known as Core, which was based on the mobile-oriented Pentium M variant of the P6 microarchitecture~\cite{hinton.talk2010}, with roots spanning back to the mid-1990s.

\paragraph{Examples of CPU Simplification.}
While it is difficult to directly compare CPUs across different microarchitectures, there are three key features that highlight some of the decisions made to simplify the Core 2 Duo's CPU cores compared to the NetBurst microarchitecture, as a motivating example.
First, the Core microarchitecture initially consisted of 14 pipeline stages, compared to the 31 pipeline stages found in the last generations of the NetBurst microarchitecture. 
Second, CPU clock frequencies were significantly lowered as well, with the first generation of Core 2 Duo desktop CPUs achieving a peak frequency of \SI{2.67}{\giga\hertz}~\cite{intel.core2duoe6700.ark.2006}, compared to a peak frequency of \SI{3.8}{\giga\hertz} for the Pentium~4 desktop CPUs~\cite{intel.p4570j.ark.2006}.
Third is the removal of multithreading support, which allows a CPU to simultaneously execute two applications on the same CPU core through strategic resource sharing. Multithreading was not included in the initial Core 2 Duo CPUs, despite being a prominent feature of the Pentium~4 architecture.
It is important, however, to note that these were not long-term changes: deeper pipelines, high clock frequencies, and multithreading support all returned in future Intel multicore CPUs, as continued manufacturing process scaling and microarchitectural innovations enabled these and other features to be implemented more efficiently.

\paragraph{Dynamic Voltage and Frequency Scaling.}
Given the significant implications of the power wall, several low-power techniques were developed around the same time as the emergence of multicore CPUs.
One such technique that has become commonplace is \emph{dynamic voltage and frequency scaling} (DVFS)~\cite{macken.isscc1990}.
With the dramatic performance increase of CPUs through the 1990s, a single CPU was capable of performing significant amounts of computation in a short amount of time. However, this peak throughput is achievable only if there is enough computation to perform.
In reality, programs frequently encounter \emph{stalls}, where the CPU is unable to maximize its throughput and can eventually sit idle as it waits for more instructions to be ready to execute.
Stalls can occur for a number of reasons, such as having limited opportunities for ILP in a highly-pipelined superscalar CPU, or having frequent memory or I/O operations that can take hundreds to thousands of cycles to service.

When the CPU is unable to achieve peak throughput, significant energy is wasted by running the CPU at its full voltage and frequency.
The dynamic power $P_{dyn}$ consumed by a CPU to execute a program can be modeled as a function of its supply voltage ($V$) and frequency ($f$):
\begin{equation}
    P_{dyn} \propto \alpha \times C \times V^2 \times f
    \label{eq:dynpower}
\end{equation}
where $\alpha$ is an activity factor, and $C$ is the switched load capacitance of the CPU's transistors.
Over time, if the CPU continues at full dynamic power during periods of underutilization, a large amount of energy is wasted.
DVFS techniques can detect when the CPU is stalling or being underutilized, and reduce the supply voltage and/or the clock frequency of the CPU.

\begin{figure}[h]
    \centering
    \includegraphics[width=0.45\linewidth]{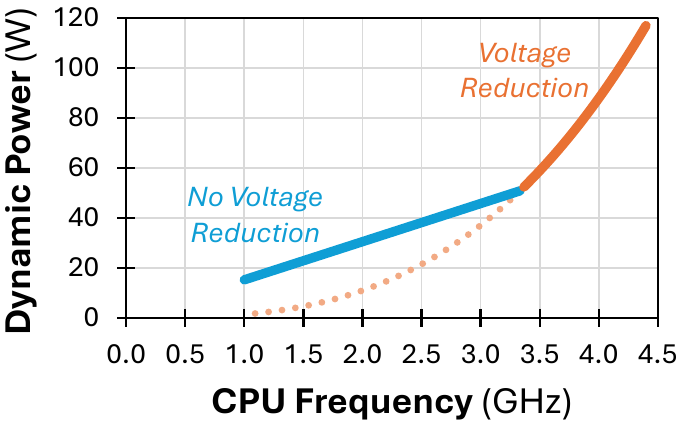}
    \caption{Dynamic power as a function of DVFS-selected frequency. Solid line shows the actual DVFS behavior, while the dotted line shows idealized further reductions to dynamic power if $V_{min}$ was not a lower limit to voltage scaling.}
    \label{fig:dvfs}
\end{figure}

One limitation of DVFS is $V_{min}$, the minimum operating voltage of the CPU core.
A lower bound on $V_{min}$ is the \emph{threshold voltage} of a transistor, which is the minimum voltage at which the transistor effectively operates as a digital switch.
The threshold voltage is a function of the manufacturing process technology.
Practically, though, $V_{min}$ is notably higher than the threshold voltage, in order to ensure the reliable operation of the CPU.
Between the nominal operating voltage ($V_{dd}$) and $V_{min}$, DVFS lowers the \emph{voltage} of the selected CPU core.
As the voltage is linearly correlated with the CPU core frequency, this results in a cubic reduction in dynamic power, based on Equation~\ref{eq:dynpower}.
Once DVFS reaches $V_{min}$, it can continue to lower the frequency, but must leave the voltage at $V_{min}$, resulting in a linear reduction in dynamic power.
Figure~\ref{fig:dvfs} shows this relationship for a hypothetical CPU, with a base frequency of \SI{4.0}{\giga\hertz} at a thermal design point (TDP) of \SI{88}{\watt}, $V_{dd}$ = \SI{1.2}{\volt}, $V_{min}$ = \SI{1.0}{\volt}, and a turbo boost frequency of \SI{4.4}{\giga\hertz}.

Given that each core in a multicore system has the ability to execute a different thread or workload from the other cores, it is inefficient to manage DVFS settings at a global level.
However, early approaches to DVFS relied on large off-chip switching voltage regulators, which each consumed significant area and power, and which had response times on the order of tens of microseconds.
Within a few years of the introduction of DVFS in commercial systems, on-chip voltage regulators were developed.
These on-chip regulators, often called \emph{buck converters}~\cite{kim.hpca2008}, are capable of performing sub-\SI{100}{\nano\second} voltage switching, at $>$90\% efficiency.
Due to their low cost and fast response, buck-converter-based DVFS is now widely used in multicore CPUs, where each CPU core has its own on-chip regulator to allow for per-core DVFS throttling.
The voltage and frequency settings for each core are typically selected in the operating system, using a \emph{governor} (see Section~\ref{sec:sw:scheduling}).

\subsection{Sharing Caches and Main Memory}
\label{sec:hw:llc}
\label{sec:hw:mem}

At a high level, general-purpose CPU design aims to accommodate a wide range of workloads by providing enough hardware resources to execute most, if not all, of the expected CPU workloads with reasonable performance and efficiency.
As a result, CPU design tends to provision resources for close to worst-case behavior, while enabling optimizations for the common case.
Multicore CPUs provide an opportunity to mitigate some of the costs of worst-case resource allocation in single-core CPUs.
For example, in the single-core model, the cache hierarchy 
should ideally provide enough cache capacity to handle common workloads that exhibit large cacheable data footprints (e.g., workloads with memory access patterns that exploit spatial and/or temporal locality across multiple megabytes of data).

In a multicore CPU, the ability to \emph{share} (i.e., pool together) resources across multiple cores can allow us to support near-worst-case behavior for a subset of cores (e.g., if one core runs a worst-case workload), while keeping \emph{total} resource allocation at a more modest level.
Returning to the cache capacity example, while a multicore CPU catering to the worst case (e.g., a situation where each core runs a large-footprint workload) could potentially require several megabytes of cache capacity \emph{per core}, multicore designers instead reduce the per-core amount of cache in the CPU, and instead allow cores to share much of their cache capacity with each other.
The two most prominent examples of resource sharing in a multicore CPU are in the memory hierarchy:
the \emph{last-level cache} (LLC) and main memory.

One issue with sharing resources across CPU cores is \emph{interference}~\cite{bitirgen.isca2008}, where due to the finite amount of available resources, multiple cores may contend with each other for their desired share of these resources.
Let us look at a contrived example, with a two-core CPU, where one core is servicing a cache miss that must retrieve data from main memory.
As the data for that miss arrives at the LLC, the LLC must evict an existing cache block (assuming that no invalid blocks remain).
Interference can occur when the eviction is for a cache block belonging to the other core, as that core will now miss on a subsequent access to the evicted block.
Had the first core not triggered the eviction, this subsequent access would have been a cache hit, avoiding the long-latency miss to main memory.
The resource contention that results from multicore interference can hurt the effective performance and energy consumption of the CPU (Section~\ref{sec:metrics} discusses metrics to quantify this impact).
As a result, several approaches have been proposed to mitigate interference in the memory hierarchy (see Section~\ref{sec:mem:interference}).

\paragraph{Multicore Cache Hierarchy.}
A multicore CPU maintains a combination of \emph{private caches} (i.e., caches accessible by only a single CPU core) and \emph{shared caches} (i.e., caches accessible by multiple CPU cores) that are made of \emph{static random-access memory} (SRAM).
For each core, a typical modern multicore CPU includes a private L1 instruction cache, a private L1 data cache, and a private L2 unified cache (i.e., a cache that holds both instructions and data).
The LLC (the L3 cache in typical modern multicore CPUs) is shared naively (i.e., without any core-based partitioning) across all cores in the CPU: any workload executing in the CPU can allocate a cache block anywhere in the shared cache.
This allows workloads with heterogeneous cache needs to co-exist symbiotically:
as an example, if one workload wants to use a large portion of the cache while another workload needs only a small portion of the cache, both workloads can execute concurrently while satisfying their needs.

In order to accommodate the needs of all of the cores in the multicore CPU, the LLC tries to reduce the potential for set contention, and the associated invalidations.
First, the LLC is significantly larger (e.g., on the order of multiple megabytes, reaching tens of megabytes for large contemporary multicore CPUs) than the typical LLC capacities from the single-core era (e.g., the L2 caches of that era, then the LLC of the CPU, were only hundreds of kilobytes in size).
Second, the LLC has notably higher associativity, with modern CPUs implementing 24-way and 32-way set-associative caches.
While this significantly increases the area of the LLC, with some multicore CPUs using as much as half of their total die area for the LLC, the increases are amortized across all of the cores (as compared to increases in the private cache area).

\paragraph{Constructing Large Caches.}
\label{sec:hw:noc}
Modern LLCs typically make use of hierarchical designs to provide the large capacities required by contemporary workloads.
A modern LLC can contain hundreds of thousands of cache sets, 
and if a CPU designer were to maintain a monolithic array for the entire LLC capacity, they would have to deal with very large wire delays, as a bitline would potentially be shared by hundreds of thousands of rows, increasing the capacitance and resistance of both the wire itself and the parasitic impacts of each attached row.
The increased capacitance and resistance translate into linear increases in delay and energy, and if left unaddressed can undermine the utility and efficiency of the LLC.
Compounding the issue is the need for more cache \emph{ports}, as multiple cores may want to access the cache at the same time, and a single-ported LLC can lead to starvation during periods of heavy cache request queuing.

To work around capacity and port limitations, modern caches rely on the concept of \emph{cache slicing}, where a cache level is decomposed into \emph{multiple} smaller pieces.\footnote{Cache slicing has been referred to by several terms over the years, including cache interleaving and cache multi-banking. Early works on cache interleaving date to 1980~\cite{smith.csur1982}, and initial implementations interleaved (i.e., distributed) the words within a single cache block across multiple cache banks. This chapter uses the term cache slicing to refer to cache block interleaving across cache slices/banks, where a single slice/bank contains the entire cache block~\cite{sohi.asplos1991}.}
Each cache slice contains a subset of the total number of cache sets in the level, and is further partitioned by the number of ways.
The slicing requires each cache access to first go through a slice decoder to identify which slice to look up.
While early slice decoder implementations used a subset of the cache block address bits (e.g., using two of the cache index bits to decide which of four slices the block is assigned to), several commercial CPUs employ complex hash-based set mapping functions (e.g., \cite{lempel.hc2011}) to avoid \emph{hotspots} (i.e., the uneven distribution of cache accesses across slices).

Inside a slice, to further manage scalability, the slice is typically partitioned into a \emph{tag store} (i.e., a tag directory, which stores metadata about each cache block in the slice) and a \emph{data store} (i.e., the actual data inside each cache block).
While tag--data partitioning is an independent concept from cache slicing, it also helps to reduce cache energy by not activating the data bitlines unless a cache hit occurs (albeit at the expense of increased latency due to serializing the tag and data lookups).
Note that for very large caches, the data store of a cache slice can be further partitioned into multiple 
two-dimensional \emph{subarrays} of SRAM~\cite{huang.jssc2013}, to further reduce power consumption.

\begin{figure}[ht]
    \centering%
    \begin{subfigure}[c]{0.48\linewidth}
        \includegraphics[width=\linewidth]{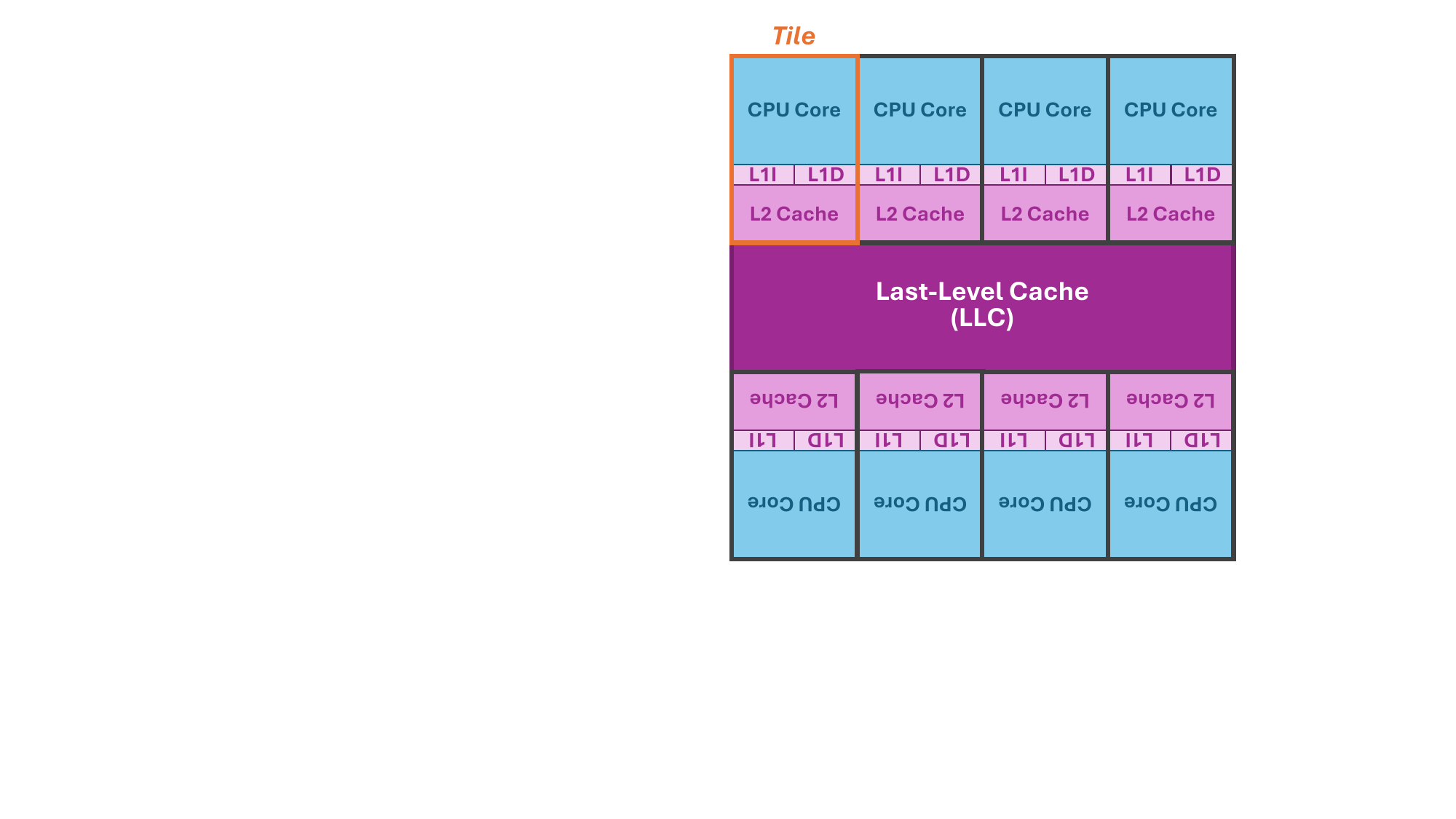}
        \subcaption{Tile without a piece of the LLC.}
        \label{fig:tile:noslice}
    \end{subfigure}%
    \hfill%
    \begin{subfigure}[c]{0.48\linewidth}
        \includegraphics[width=\linewidth]{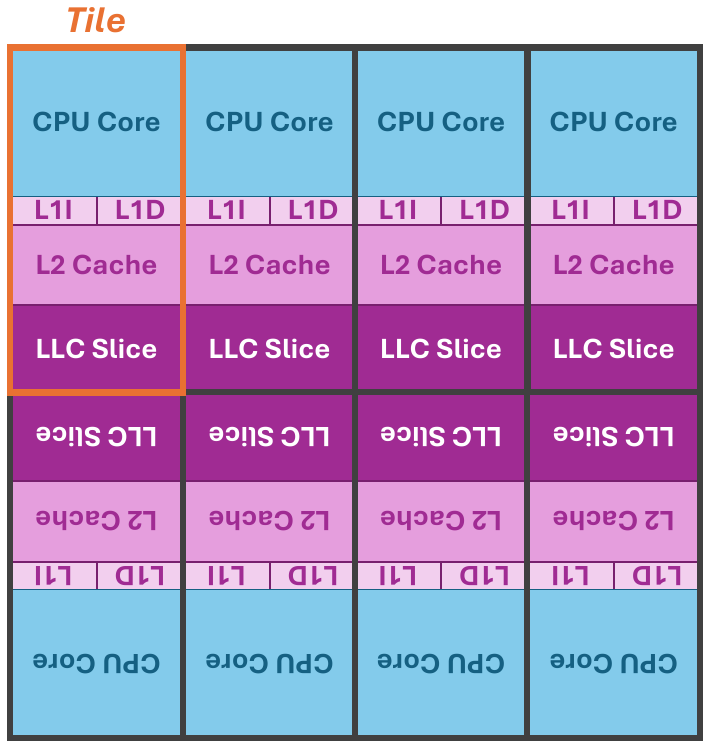}
        \subcaption{Tile containing an LLC slice.}
        \label{fig:tile:slice}
    \end{subfigure}%
    \caption{Two approaches to tiling a multicore CPU layout. Note that tiles on the bottom row are intentionally shown upside down to represent the stamping of identical tiles.}%
    \label{fig:tile}%
\end{figure}

There are two general approaches to laying out the shared LLC on a multicore CPU chip, both of which support cache slicing.
As mentioned in Section~\ref{sec:hw:cores}, a multicore CPU is made up of multiple tiles, where a tile includes a CPU core and its private caches.
The first approach for LLC layout allocates a fixed amount of cache outside of these tiles, as shown in Figure~\ref{fig:tile:noslice}, using a separate tile for the LLC.
Often with this layout, the controller for the LLC is centralized, and each core has access to the cache controller via a \emph{bus} (i.e., a shared wire across all cores), though some implementations maintain an independent controller for each cache slice, in which case a \emph{crossbar} is used.
\todo[figure showing interconnects]
The second approach for LLC layout includes a slice of the LLC with each CPU core tile, as shown in Figure~\ref{fig:tile:slice} (leading to a one-to-one correspondence between core count and cache slice count).
With this layout, each cache slice typically maintains an independent controller, and the controllers are connected together using a \emph{ring interconnect}~\cite{lempel.hc2011, huang.jssc2013}.

A key challenge with large LLCs is access time.
Conventionally, monolithic caches provide \emph{uniform cache access}, where all cores can access any block in the cache with the same latency (assuming no interference).
As LLCs have become larger and sliced, cores often deal with \emph{non-uniform cache access} (NUCA)~\cite{kim.asplos2002}, where it can take a longer time for a core to access a remote slice (i.e., the additional time required to traverse the interconnect) than it does to access its local slice.
A NUCA cache can incur higher latencies than a cache with uniform cache accesses, but the slicing enables additional bandwidth by allowing concurrent accesses to each slice, mitigating some of the performance loss.

\paragraph{Main Memory.}
Main memory modules (which consist of \emph{dynamic random-access memory}, or DRAM, chips in modern systems) are connected to a CPU by a \emph{memory channel} (see Section~\ref{sec:mem} for details).
Any core in a multicore CPU can access any of the CPU's memory channels, by accessing the on-chip \emph{memory controller} that corresponds to the desired memory channel.
As is the case for single-core CPUs, the memory controllers in a multicore CPU receive requests from the LLC.
Given the sliced architecture of the LLC, each slice typically has the independent ability to dispatch cache misses to a memory controller.
This is achieved by providing each slice with dedicated \emph{miss status holding registers} (MSHRs)~\cite{kroft.isca1981}, which make use of fully-associative structures built using content-addressable memories (CAMs) to track in-flight cache requests.
As there is not usually a one-to-one correspondence between cache slices and memory controllers (the typical multicore CPU has more cache slices than memory controllers), a slice can dispatch misses to any controller, and chooses the correct controller based on how the system partitions memory addresses across memory channels.

For a given mix of workloads, a multicore CPU with \emph{n}~cores can potentially issue cache misses to main memory at as much as \emph{n}~times the rate that a single-core CPU does (note that this assumes similar core architectures in both CPUs, and depends heavily on the specific workloads and on the impact of cache interference between cores).
This requires the main memory to scale up its ability to respond to these cache misses, and modern memory subsystems have two examples of this.
First, more MSHRs are provided at the LLC in order to sustain a greater number of concurrent memory accesses.
An LLC miss must stall any time the cache does not have a free MSHR available to allocate.
If there are not enough MSHRs in the system, then such stalls are more likely to occur.
As each cache slice often has its own set of MSHRs, multicore CPUs take advantage of the increased slice count to provide more MSHRs at a low cost, compared to scaling up a monolithic CAM.
Second, the memory bandwidth has been scaling up to accommodate the increased demand.
The memory bandwidth provided by DRAM has been increasing significantly over the last two decades, through a combination of
(1)~increasing bus frequencies for the memory channel and
(2)~the emergence of new DRAM types such as DDR5~\cite{ddr5.spec} and High-Bandwidth Memory (HBM)~\cite{hbm.spec}.
While memory bandwidth can also be scaled by increasing the number of memory channels per CPU, multicore CPUs have largely avoided this approach due to the limited number of pins available in the CPU chip package.

The sharing of a single main memory across multiple cores can introduce new changes to the behavior of three properties, compared to their behavior with single-core CPUs.
First, the optimal choice of an \emph{address interleaving} scheme (i.e., which bits index which memory structures) can change depending on the type of workload executing on the CPU.
Second, the optimal choice of \emph{row policy} (i.e., whether a DRAM row is left open after all currently-queued requests are serviced) can depend on the type of workload as well.
Third, the \emph{memory scheduling algorithm} (i.e., the order in which queued DRAM requests are serviced) may need to change to avoid starving requests in certain scenarios.
Section~\ref{sec:mem:policies} discusses each of these in detail.

\subsection{Coordinating Memory Requests Across Cores}
\label{sec:hw:coord}

One key artifact of the memory hierarchy of a multicore CPU is that cores do not always write updates to globally-visible locations.
For example, in a typical multi-level hierarchy that employs write-back caches, a core will write data to its private L1 data cache, but this updated value will not be immediately visible to other cores in the CPU, as they cannot directly access another core's private cache.
This is at odds with the architectural model (i.e., abstraction) that is presented to programmers, where the cores have access to a global \emph{shared memory} (which is not the same as main memory; see Section~\ref{sec:mem:shared} for more details).
As part of this shared-memory model, a data update written by a core to shared memory will become visible to other cores.
In contrast, caches are a \emph{microarchitectural} optimization that is not, in principle, exposed to programmers as part of the architectural abstraction.\footnote{This is in part because (1)~caches are designed to be hardware-managed structures that operate transparently to the programmer, and (2)~cache configurations can differ between different CPU models belonging to the same instruction set architecture. In practice, programmers use knowledge of the design of the cache hierarchy to optimize program performance, while still expecting the behavior of the system to adhere to the semantics of the shared-memory programming model.}
As such, any interaction that a core has with a cache must \emph{appear} to the program as if it's taking place globally in shared memory, as this is what a programmer expects.

In order to maintain program correctness, multicore CPU must ensure that data updates are coordinated across all of the cores (as is the case in any parallel computer architecture).
For the shared-memory programming model typically used for multicore CPUs, this involves two types of coordination~\cite{nagarajan.book2020}:
(1)~\emph{cache coherence}, which ensures that updates to a \emph{single} unit of data (e.g., one cache block) are made visible to and ordered across all cores; and
(2)~\emph{memory consistency}, which ensures that updates across \emph{multiple} units of data are interleaved according to a predetermined policy across all cores.
Sections~\ref{sec:mem:coherence} and \ref{sec:mem:consistency} discuss cache coherence and memory consistency, respectively, in more detail.
\todo[make cores vs. threads consistent]

\subsection{Scaling to Many Cores}
\label{sec:hw:manycore}
\label{sec:manycore}

While the term \emph{multicore} does not in theory place any limits on the number of cores in a CPU, there is a distinction made between multicore CPUs with smaller core counts (e.g., under 24~cores in contemporary systems) and \emph{manycore CPUs}, which are multicore CPUs that consist of several dozens of cores.
This distinction is made because of key scalability challenges that become prominent as the core count increases significantly.
For the smaller core counts, the interconnects described in Section~\ref{sec:hw:noc} can enable reasonable parallel performance.
However, at the manycore level, the significant increase in contention makes both bus-based and ring-based interconnects infeasible.
As a result, specialized research and development has focused on how to provide more scalable communication at high core counts.

The initial ideas that would evolve into manycore CPU design stem from the Raw microprocessor project at MIT~\cite{waingold.computer1997}.
Conceived around the same time as the Hydra CMP (Section~\ref{sec:motivation:history}), the Raw CPU took a more extreme approach to CPU core simplification, arguing that sophisticated compilers could offload the need for complex ILP mechanisms.
As a result, the Raw CPU consisted of multiple small \emph{tiles}, where each tile included a very simple CPU core along with a small piece of cache.
The tiles were meant to be composable: depending on the needs of a platform, and on the available transistor count, a manufacturer could stamp out more or fewer tiles depending on their needs.
By making each tile small, the distance between tiles (and, thus, the distance between cores) and intra-tile wire lengths would both be short, allowing for the CPU to run at a faster clock frequency without excessive power consumption.
To enable composability with varying tile counts, the tiles connected to each other using a packetized \emph{mesh network} (a two-dimensional interconnect), including configurable on-chip routers.
The first Raw CPU, prototyped in 2002, included 16~tiles on a single chip.

Unfortunately, there is no precise definition of a core count or of specific properties that a manycore CPU must have, and the perceived distinction between manycore CPUs and more conventional multicore CPUs has shifted as capabilities have evolved over the last two decades.
Aside from mesh-based tile organizations, other properties observed in several manycore CPUs include non-uniform cache access (NUCA) architectures with multiple cache slices, and the replacement of hardware cache coherence with software-driven message passing interfaces.
While not a comprehensive list, examples of commercial CPUs that have been considered to be manycore include the Tilera TILE~64 (a commercialization of the MIT Raw CPU), the Intel Xeon Phi series, and various CPUs from PEZY, including their 2048-core PEZY-SC2 released in 2017.

%% file: sections/memory.tex
\section{Managing Memory}
\label{sec:mem}

The cores in a multicore CPU share memory with each other, as discussed in Section~\ref{sec:hw:mem}.
This sharing introduces a number of issues that must be considered by both programmers and architects.
This section will discuss several of these issues.
To start, it defines what it means to share memory, both from a software perspective and a hardware perspective.
Then, it discusses main memory management, and how management policies in hardware can impact the performance of threads running on a multicore CPU.
Finally, it discusses examples of cache coherence and memory consistency protocols that ensure correct program behavior in a shared-memory environment.

\begin{figure}[!h]
    \centering
    \includegraphics[width=0.8\linewidth]{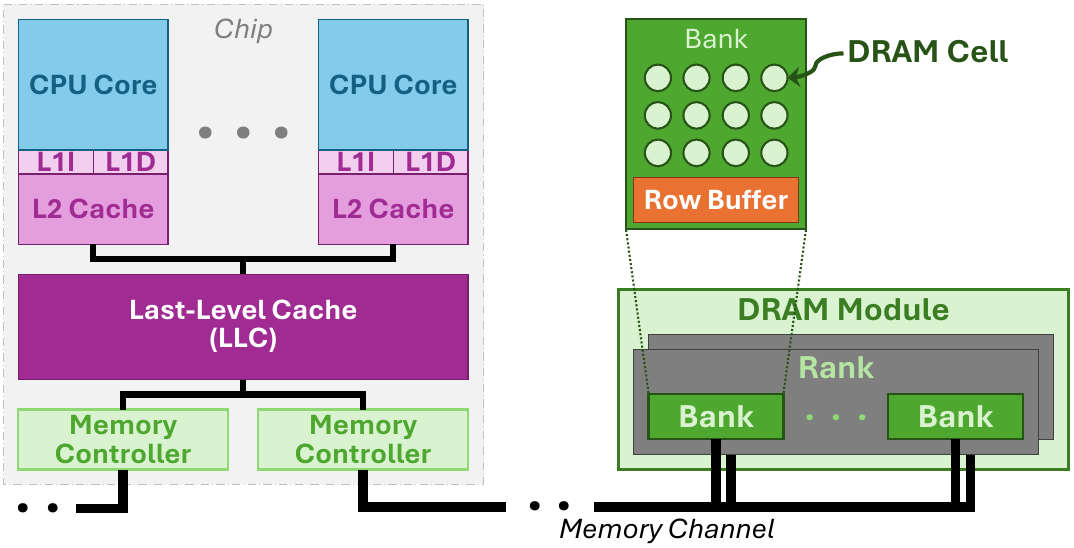}
    \caption{Example memory hierarchy for a multicore CPU, showing shared on-chip last-level cache and shared off-chip main memory.}
    \label{fig:memory}
\end{figure}

To guide the discussion in this section, Figure~\ref{fig:memory} shows an example of how the hardware in a memory subsystem is shared across the cores in a multicore CPU.
All of the cores in a multicore CPU share a last-level cache (LLC; see Section~\ref{sec:hw:mem} for LLC design details).
If a cache request misses in the LLC, it must go to main memory, which is made up of DRAM.
The physical address space of a system enables each byte physically available within the main memory to be accessed using a unique address.
The physical memory is split up across one or more \emph{memory channels}, where each channel connects to one or more DRAM \emph{modules}.
Each channel is managed independently of one another, and access management and maintenance tasks (e.g., refresh) are handled by a dedicated \emph{memory controller} for the channel (see Section~\ref{sec:mem:policies}).
Each module contains a series of DRAM chips, which are grouped into one or more \emph{ranks}.
All the chips belonging to a rank operate in lockstep (i.e., they always perform the same operations on the same row/column of the same arrays in each chip).
A rank consists of several \emph{banks} of memory, with each bank physically striped across all of the chips in a rank.
Logically, a bank operates as a single two-dimensional array of DRAM cells, where a cell consists of a capacitor and a transistor, and can hold one bit of data.
Due to the small charge capacity of the capacitor, the memory controller cannot directly perform data operations on the DRAM cell, and instead loads (i.e., activates) one row from a bank at a time into the \emph{row buffer}, from which the controller can issue reads from and writes to the activated row.

\subsection{Shared-Memory Model}
\label{sec:mem:shared}

From the first symmetric multiprocessor computer, the Burroughs D825~\cite{anderson.fjcc1962}, the majority of multiprocessor systems have enabled CPU cores to share the physical main memory with each other.
Through the decades, to mitigate the long latencies of memory, several microarchitectural enhancements (e.g., caches, memory speculation) have extended the memory subsystem hardware, but have done so transparently to the programmer.
From the programmer's perspective, multicore CPUs predominantly make use of a \emph{shared-memory model}, which at a high level resembles primitive computers.
In the shared-memory model, each core has access to a single shared physical memory (which may be implemented in a centralized or distributed manner).
When any core performs a store to an address in this single shared memory, the update becomes visible to all cores that subsequently load from that address.
While in principle the shared-memory model sounds simple, it has two key implications on the design of multiprocessor systems (and, by extension, multicore CPUs).

First, for a core, one needs to define when a store is considered to be performed (i.e., the moment at which the stored value becomes visible to other cores).
For example, for an out-of-order core, a popular definition is that a store is performed once the store instruction is committed.
Prior to commit, there are several events that can cause an executed store to be squashed (e.g., exception, misspeculation), making the commit event the earliest time that the store is guaranteed to safely proceed.
As a result, stores are buffered inside the core, and the store operation to the first-level cache is not initiated until after the store instruction commits.
A physical \emph{store buffer} (sometimes called a \emph{write buffer}) contains all committed writes that are yet to be issued to the cache.
Depending on the memory consistency model, this store buffer may or may not be considered a part of shared memory; if it is, the store values in the buffer must be made available to other cores in a way that does not violate the consistency model (see Section~\ref{sec:mem:consistency} for more).

Second, the CPU must ensure that microarchitectural optimizations for memory operations obey a consistent set of rules, which are part of the instruction set architecture (ISA), so that programmers do not encounter program correctness issues during execution.
All private and shared caches are considered part of the shared memory, which means that cache coherence is needed to propagate the value of a cache store to other caches and cores, as Section~\ref{sec:mem:coherence} discusses.
Both within a thread and across threads, the interleaving of a store with other stores and with loads must obey a set of rules that are promised to the programmer, which are defined as part of the memory consistency model.
As Section~\ref{sec:mem:consistency} discusses, some memory consistency models have strict interleaving expectations, which can simplify programming complexity at the expense of high performance overheads, while more relaxed memory consistency models can allow different cores to observe different interleavings (following a defined set of guarantees) in order to improve performance.

\subsection{Main Memory Policies}
\label{sec:mem:policies}

As discussed in Section~\ref{sec:hw:mem}, there are three types of policies for main memory management that are impacted by the introduction of multiple cores:
address interleaving, row policy, and memory scheduling.
Each of these policies is implemented inside the memory controller.
Unlike the single-core case, where the choice of policy for each type can be determined by analyzing program behavior, the non-determinism that exists for shared-memory interactions in a multicore CPU make it difficult to easily choose a single optimal policy.

\paragraph{Address Interleaving.}
Address interleaving (also known as memory interleaving) schemes~\cite{balasubramonian.book2019} determine the way in which address bits are assigned to index different levels of the main memory organization, in an attempt to take advantage of spatial locality and memory-level parallelism (MLP)~\cite{glew.asploswaci1998}.
As shown in Figure~\ref{fig:memory}, a typical DRAM-based main memory includes one or more memory channels, where each channel contains one or more DRAM modules.
Within each module, there are multiple chips, with the chips grouped into one or more ranks.
Each rank (a group of DRAM chips that operate in lockstep) contains multiple banks of DRAM, where each bank can mostly operate in parallel with each other.
Inside each bank, at most one row of DRAM can be open at a time, where a row holds multiple columns of data (with each row in commonplace DRAM types containing a few kilobytes of data).

\begin{figure}[!h]
    \centering
    \includegraphics[width=\linewidth]{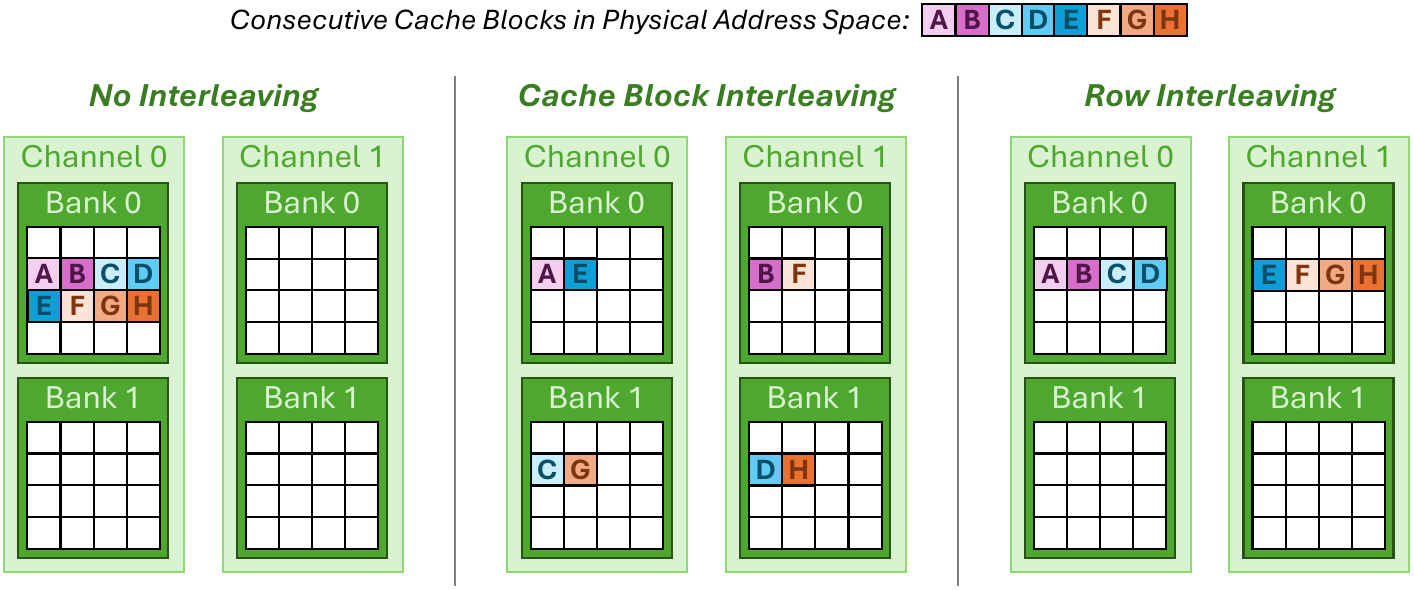}
    \caption{Example of how address interleaving schemes impact the placement of cache blocks in main memory, for a series of cache blocks that are consecutive in the physical address space. A specific interleaving scheme is chosen by determining which bits of the physical memory address are used to index the different levels of the memory hierarchy (in this example, the channel, the bank, the row, and the column).}
    \label{fig:addrinterleaving}
\end{figure}

Figure~\ref{fig:addrinterleaving} illustrates two common interleaving schemes that optimize locality and MLP over a basic approach without interleaving. 
In the first scheme, \emph{cache block interleaving}, consecutive cache blocks in the memory address space are distributed to different banks (and in some cases to different memory channels) to allow requests to the two blocks to be serviced concurrently.
In the second scheme, \emph{row interleaving}, consecutive cache blocks stay in the same row to maximize hits to the already-open row, but consecutive rows are distributed to different banks/channels.
While cache block interleaving is a popular scheme for single-core CPUs, it can introduce issues in a multicore CPU.
For example, with a multiprogrammed workload (i.e., when multiple cores are executing threads that belong to different processes), one thread may tie up many channels at the same time by accessing several consecutive cache blocks.
This would increase the chance of interference for any of the other threads trying to access data in memory, by increasing the likelihood that the channel will be busy.
Worse off, if two different processes access several consecutive cache blocks, this can generate a large number of \emph{bank conflicts} (i.e., \emph{row conflicts}) in DRAM that could have been avoided with a row interleaving scheme.
A non-interleaved memory (i.e., where consecutive rows map to the same bank) can reduce the probability of bank conflicts due to interference, but at the expense of sacrificing memory-level parallelism for memory-intensive applications.

\paragraph{Row Policy.}
The row policy~\cite{balasubramonian.book2019} determines whether a row in a bank should remain open once the memory controller services all currently-queued requests for that bank.
In an open-row policy, the row stays open in case future memory requests also access the same row (due to spatial locality), avoiding the latency of re-opening (i.e., reactivating) the row when the next request to the bank arrives.
In a closed-row policy, the row is closed as soon as the currently-queued requests for that bank are completed, under the assumption that new requests are likely to access a different row, thus avoiding the row closing (i.e., precharging) latency when the next request to the bank arrives.
Variations of these policies exist, such as timeout policies that close the row a given number of cycles after the last queued request is serviced.
In a multicore CPU, the optimal policy choice can depend on the workload executing on the CPU.
For example, with a multiprogrammed workload, the likelihood of the next access to the bank being to the same row decreases compared to a single-core CPU, due to interference and competition across different processes.
However, with a multiprogrammed workload, the likelihood can potentially increase instead, as multiple threads belonging to the same process may each access data in the same row, depending on how data accesses are partitioned across threads.

\begin{figure}[!h]
    \centering
    \includegraphics[width=\linewidth]{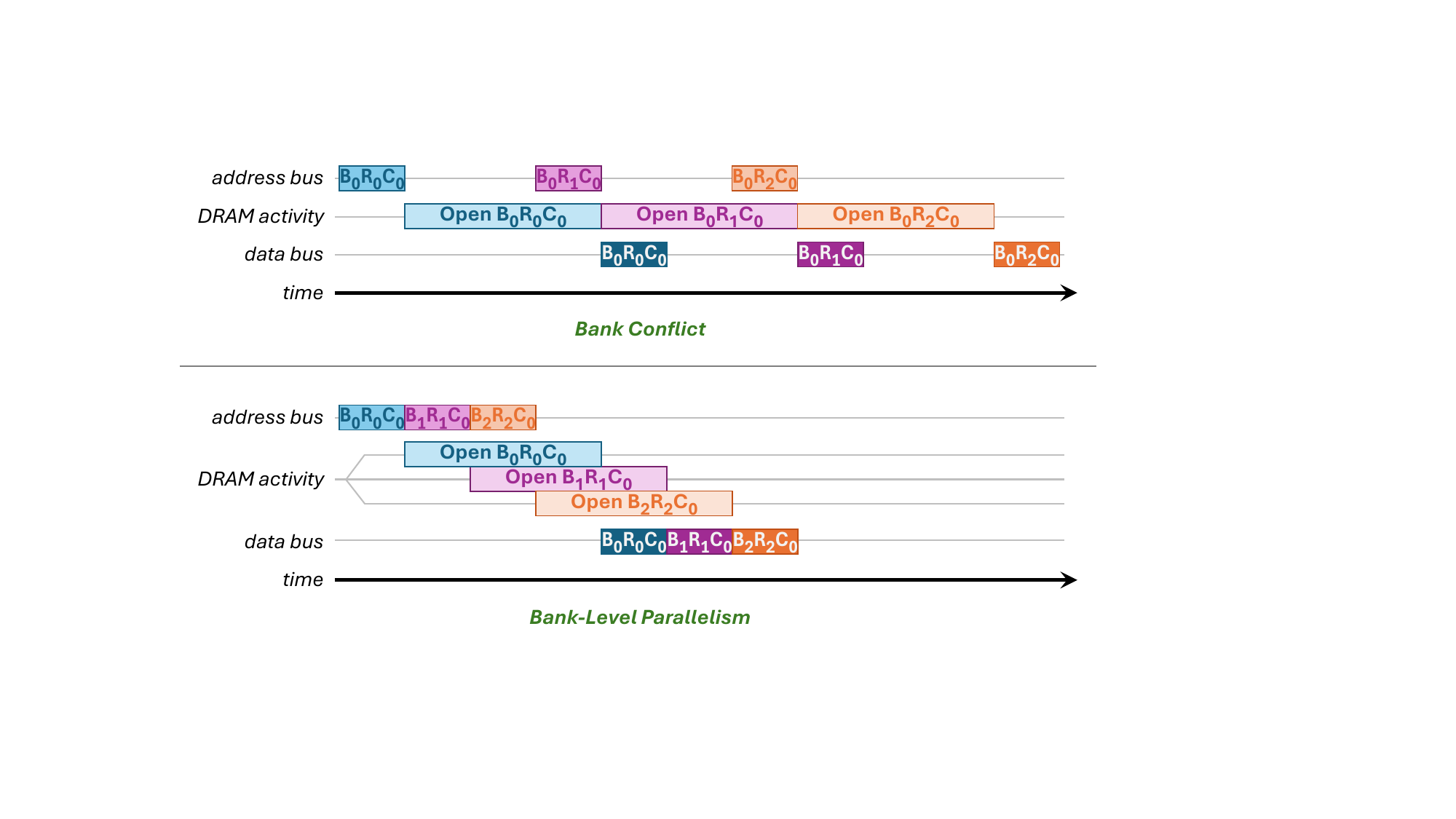}
    \caption{Scheduling of multiple read requests when a bank conflict occurs (top), and when there is bank-level parallelism that avoids a bank conflict (bottom). The notation $B_b R_r C_c$ indicates a memory access to Bank~$b$, Row~$r$, Column~$c$. This example assumes that all requests access the same memory channel. The memory controller for the channel sends a request to the memory via the address, bus just before it is safe to initiate the operation inside memory (e.g., opening a row). The controller must ensure that operations to the same bank do not overlap \emph{and} data transfers from memory to the controller on the shared data bus do not overlap.}
    \label{fig:blp}
\end{figure}

\paragraph{Memory Scheduling.}
The scheduling algorithm~\cite{balasubramonian.book2019} determines which of the pending memory requests to service next, which impacts the rows and banks that are currently active in DRAM.
As mentioned above, a DRAM read or write cannot be directly performed on the DRAM cells, and can only perform the reads or writes on data in the row buffer of a bank (which practically limits a bank to servicing at most one request at a time).
This results in significant complexity for scheduling memory requests, as the controller must consider many factors such as the time a request has been waiting for, whether the target bank for each queued request is idle or is in the middle of servicing another request, and whether the target bank has the target row currently open (i.e., activated).
Furthermore, while a memory controller can take advantage of \emph{bank-level parallelism} (BLP) to service requests to \emph{different} banks concurrently, the physical bus wires of the memory channel are shared across all banks belonging to the channel, as shown in Figure~\ref{fig:memory}.
As a result, the scheduler must also stagger requests in a way that ensures exclusive access to the memory channel bus for only one bank, when the bank needs to send data to or receive data from the controller.
Figure~\ref{fig:blp} shows how this staggering of memory requests can be coordinated for two examples:
(1)~when multiple requests target different rows in \emph{the same bank} (i.e., a \emph{bank conflict}), and must serialize the opening of each row; and
(2)~when multiple requests target different rows in \emph{different banks}, which can exploit BLP to overlap multiple row openings, and must serialize only the actual data transfer on the shared memory channel's data bus.

Ultimately, a scheduling algorithm enforces these many constraints by prioritizing currently-queued requests based on a predefined policy and on the current state of the DRAM, and by enforcing a series of \emph{timing parameters} that are declared as part of the specification of each DRAM type (e.g., DDR5~\cite{ddr5.spec}, HBM~\cite{hbm.spec}).
A commonly-implemented algorithm is known as first-ready, first-come first-serve (FR-FCFS)~\cite{rixner.isca2000}, which prioritizes requests to already-activated rows over older requests.
FR-FCFS does this to improve \emph{average memory access time} (AMAT), as the time required to activate a row, and to precharge (i.e., close) the row after requests are finished, can both take as long as the read and write requests themselves~\cite{ddr5.spec}.
Memory controllers in commercial multicore CPUs continue to use FR-FCFS without introducing any notion of thread or application awareness~\cite{balasubramonian.book2019}, which can exacerbate interference across threads (see Section~\ref{sec:mem:interference}).

\subsection{Mitigating Interference}
\label{sec:mem:interference}

The combination of private caches and shared caches attempts to balance the impact of interference with the need to provision resources for worst-case behavior:
private caches allow each CPU core to manage a small working set of its data without first-order impacts of interference, 
while the shared cache enables resource pooling to reduce the frequency of long-latency misses to main memory for workloads with large data footprints.
However, it is still possible for memory interference, at the caches and/or at main memory, to negatively impact one or more cores.
This section looks at three such examples, as well as potential mechanisms to mitigate them.

Our first example examines how cache requests from one core can impact the management of a private cache belonging to a different core. If the shared LLC is inclusive of upper-level private caches, and one core evicts a cache block from the LLC belonging to a second core, the second core will also have to evict that cache block from its private caches \emph{even though the private cache space is not being used by the first core}.
To address this (as well as pollution from hardware prefetchers), some multicore CPUs employ a \emph{non-inclusive} or \emph{exclusive} policy for the LLC or for the last private cache level~\cite{backes.memsys2019}.
On a miss to main memory, the non-inclusive policy acts like the inclusive policy, where a copy of the data is placed in the LLC as well as in the private cache(s).
When a cache block eviction takes place at the LLC, the non-inclusive policy acts like the exclusive policy, where the eviction does \emph{not} trigger an eviction in the upper-level private caches.

Our second example examines how shared cache utilization by one core can impact the available shared cache capacity, and thus the performance, for a different core.
In a conventional multicore CPU, shared resources are left \emph{unmanaged}, i.e., there is no active mechanism to enforce that each core receives a fair portion of the resources.
This unmanaged approach is often useful to allow the easy distribution of these resources across cores based on the heterogeneous needs of the threads that each core is executing (e.g., in a two-core CPU, one core runs a thread that has a large \emph{working set} and uses up most of the LLC, while the other core runs a thread that has a small working set and can make do with whatever capacity the first thread does not use).
However, when the heterogeneous needs are more extremely unbalanced, one or more of the threads may incur significant \emph{slowdowns}.
In our example, if one greedy thread is using most of the LLC, cache blocks belonging to other threads may constantly be evicted, hurting those threads' performance significantly.
To address this, several works have proposed \emph{cache partitioning}, where some or all of the ways and/or sets of a cache are assigned to a specific core or thread.
Strict cache partitioning could ensure that the cache blocks from the other threads in our example do not get evicted, as the greedy thread would not be allowed to use partitions belonging to other threads.
One commercial implementation of cache partitioning is Intel's Cache Allocation Technology~\cite{herdrich.hpca2016}.

Our third example examines how memory scheduling algorithms can introduce unintentional slowdown for threads.
Recall from Section~\ref{sec:mem:policies} that the commonly-used FR-FCFS scheduling algorithm does not account for any sort of thread or application awareness, and solely focuses on reducing the number of row activations and precharge operations.
While reducing activate and precharge operations help decrease the AMAT for a thread that can exploit row locality, they can unfairly slow down the other threads in a multicore CPU, by continuing to prioritize requests from the thread whose row is already activated.
In this example, one thread is generating many loads and stores to the same DRAM row, which is currently activated, while the other threads each have a single load that is waiting to access a different row in the same bank.
FR-FCFS will look at all of the queued requests, and prioritize the requests from the first thread because their target row is already activated.
This causes all of the requests from the other threads to wait longer in the queue, and in extreme cases can lead to unintentional starvation for these threads.
To address this, researchers have proposed a number of memory schedulers that explicitly augment the memory request metadata with a thread ID, and incorporate information about the thread into the scheduling algorithms.
As one example, the memory controller can use lightweight runtime metrics to predict which threads are being slowed down due to memory interference, and can prioritize requests from these threads before prioritizing row locality~\cite{mutlu.micro2006}.

\subsection{Cache Coherence}
\label{sec:mem:coherence}

As mentioned in Section~\ref{sec:hw:llc}, each core in a multicore CPU has private caches in addition to a shared cache.
Similar to conventional single-core CPUs, these caches are typically configured to be \emph{write-back caches}.
A write-back cache buffers updates to a cache block, by storing the updates in the cache and marking the cache block as \emph{dirty}.
In a single-core CPU, these updates are not made visible to lower cache/memory levels until the cache block is evicted from the current cache level.
Upon eviction of a cache block, if the block is marked as dirty, its contents are written to the next lower level to preserve the most recent version of the data; otherwise, the block can simply be dropped, as at least one lower level of the memory hierarchy also has the most recent version of the data.
For cache blocks that are written to frequently, write-back caches reduce the amount of traffic these writes induce on lower levels of the cache hierarchy.
A common alternative, write-through caches, store the update in the cache \emph{and} send every update to the next lower level.
However, write-through caches are not commonly employed in multicore CPUs due to the impact of the increased write traffic on interference.

\paragraph{The Need for Coherency With Write-Back Caches.}
In a single-core CPU, write-back caches ensure that the CPU core sees the most recent version of the data for each cache block, as the core always starts memory lookups from the top cache level in the memory hierarchy.
However, with multicore CPUs, a dirty cache block stored in one core's private cache is not immediately visible to other cores in the CPU.
This is because if the same cache lookup procedure for single-core CPUs is used, a multicore CPU's core would access its own private caches and the shared last-level cache, and \emph{not} the private caches of other cores.
If left unmodified, this can lead to correctness issues, as one core may ignore and potentially incorrectly overwrite the updates from another core.

To address the lack of visibility, multicore CPUs employ hardware mechanisms to support cache coherence.
Cache coherence techniques were initially developed for parallel systems with multiple single-core CPUs, but the same basic techniques have become applied for multicore CPUs.
At a high level, a cache coherence mechanism ensures the following rules:
(1)~a write operation to a memory address by a processor must (eventually) become visible to \emph{all} processors; and
(2)~if two processors write to the same memory address, \emph{all} processors will observe the writes in the same order.
This enforcement is typically achieved by guaranteeing two properties:
(1)~at any given time, at most one core is granted permissions to write to a memory address by the coherence mechanism (note that these are different than the write permissions maintained by virtual memory); and
(2)~when a core has been granted write permissions to a memory address, any copies of that address held in other private caches are marked as stale, and cannot be used again without first synchronizing data updates from the writing core.
There are multiple \emph{cache coherence protocols} that enforce these properties in different ways,
and a multicore CPU will typically implement one of these protocols as a fixed-function mechanism in hardware.
Note that cache coherence mechanisms are fully \emph{microarchitectural}: as caches are part of the microarchitecture and are (in theory) transparent to the programmer, the programmer expects the machine to obey shared-memory model semantics, and a value written to a specific memory address should be visible to \emph{all cores} as soon as the write occurs.

Logically, a cache coherence protocol maintains a \emph{coherence state} for each core for every memory address.
The coherence state dictates the current permissions that the protocol has granted a core for a particular memory address.
As this would require a large amount of data to be stored on chip, practical implementations of cache coherence make the following two optimizations.
First, coherence states are maintained at a \emph{cache block granularity}, and not per byte (i.e., per memory address).
One artifact of block-based coherence management is \emph{false sharing}~\cite{torrellas.icpp1990}.
If two CPU cores are accessing two different memory addresses that belong to the same cache block, the cache coherence protocol treats this as data that is shared by both cores, even though their accesses are to different pieces of data.
In the case where both cores are trying to write to these pieces of data simultaneously, the cache block will ping-pong (i.e., move back and forth) between the two cores, with one core invoking an invalidation for the other core any time it wants to write, and vice versa.
False sharing can be avoided only by ensuring that the two pieces of data are mapped to different cache blocks (e.g., by having the programmer pad data structure sizes to use up exact multiples of the cache block size).
Second, if a cache block is not currently held by a core in its private caches, the cache block is assumed to be in a coherence state designated as \emph{invalid} for that core (i.e., the core cannot currently read from or write to that cache block).

\paragraph{Exchanging Coherence Messages Between Cores.}
The cache coherence protocol is typically triggered when a core wants to change its coherence state for a cache block, which can happen for one of the three following actions:
(1)~the core accesses a cache block that it does not currently have in its private caches,
(2)~the core wants to change its current permissions under the coherence mechanism for the cache block, or
(3)~the core evicts the cache block from its private caches.
If the rules of the protocol require a state change for the action, this generates a coherence message.
The specific sequencing of messages is protocol-dependent, but typically involves sending state upgrade/downgrade requests, and receiving acknowledgments and potentially updated data values.
There are two approaches to sending messages to a destination core: snoopy coherence and directory-based coherence.

In \emph{snoopy} cache coherence~\cite{goodman.isca1983}, cores share a bus, and each core maintains its own coherence state metadata.
Each core reads \emph{all} cache coherence messages transmitted on the bus (i.e., it is snooping on all messages) to see if it needs to react to the message (e.g., if it needs to write back any dirty changes to a cache block, if it needs to invalidate the block).
For example, if a core wants to write to cache block~\emph{x}, it broadcasts a coherence message on the bus declaring its intent to acquire write permissions.
As only one core can have write permissions, every other core on the bus will see the coherence message, and if a core has a copy of cache block~\emph{x}, it will invalidate the block in all of its private caches.
The other cores require some form of acknowledgment mechanism to notify the requesting core that it has completed any necessary actions, or if it has an up-to-date version of the data.
Snoopy cache coherence is the simpler of the two approaches to implement, and works effectively when the CPU has a relatively low number of CPU cores, but the broadcast-based bus scales poorly as the core count increases, both due to the increased bus latency/energy and increased contention due to message serialization on the bus.

\emph{Directory-based} cache coherence~\cite{censier.tc1978} avoids the poor scaling of broadcasting by instead storing the coherence state metadata in a directory.
When a core triggers a coherence message, the message first goes to the directory, which stores the state of the cache block in all cores serviced by the directory.
One example implementation is to maintain a bitvector for each cache block currently held by \emph{any} of the cores, where bit~\emph{i} in the bitvector indicates whether core~\emph{i} currently has a copy of the cache block.
When the directory receives a coherence message, it looks up the metadata for the requested block, and then dispatches follow-up messages to only those cores that currently hold the cache block.
While directory-based coherence is more complex to implement, it exhibits scalability over snoopy coherence in multiple dimensions.
First, as mentioned above, coherence messages need to be transmitted to only the cores that currently hold the data, and can be implemented using more scalable interconnection networks than a bus.
Second, the directory can be partitioned into multiple slices, with each slice responsible for a subset of memory addresses.
Third, a CPU can implement multiple levels of directories, creating a directory hierarchy that further reduces coherence message traffic and metadata storage requirements.
This improved scalability makes directory-based coherence a good fit when there are a large number of cores in the CPU.

\begin{figure}[!h]
    \centering
    \includegraphics[width=0.65\linewidth]{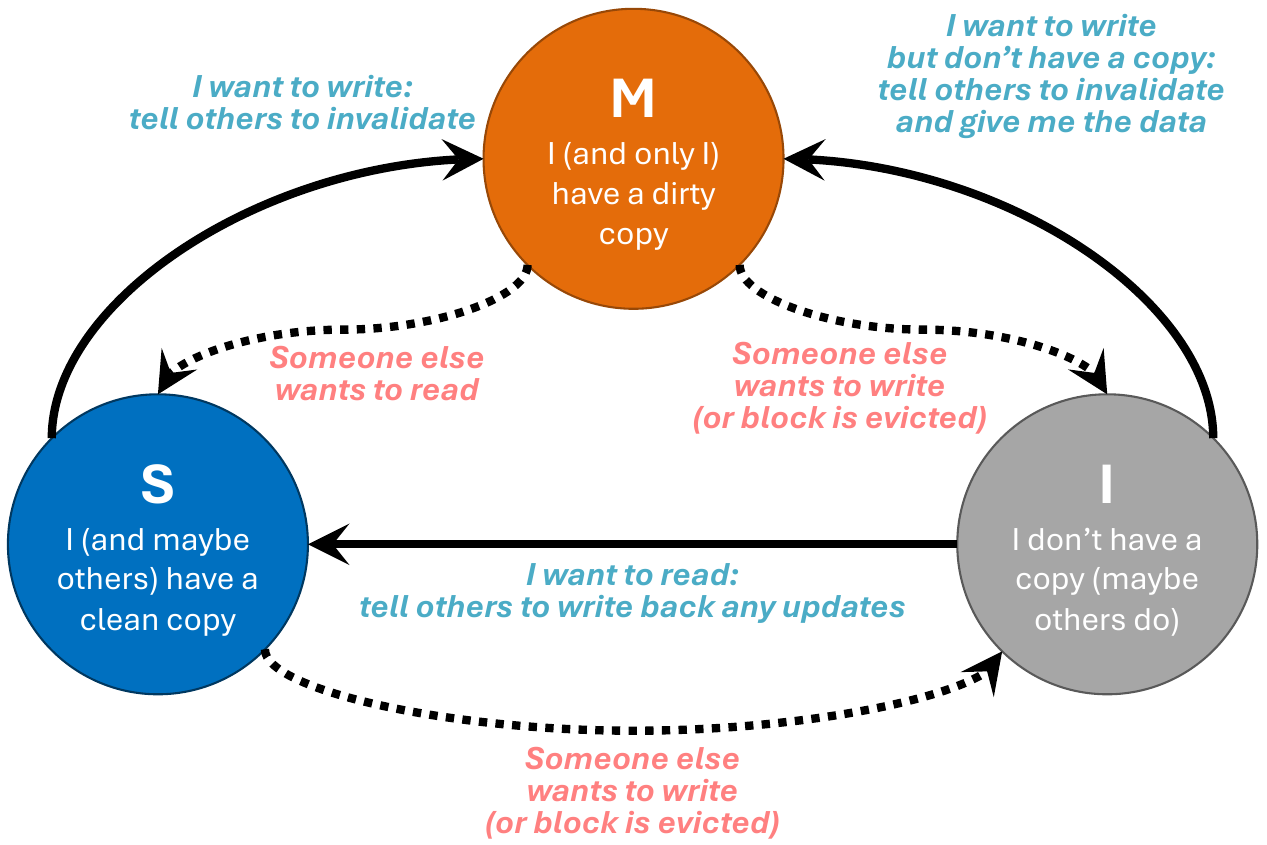}
    \caption{MSI protocol: solid lines are upgrades, dotted lines are downgrades.}
    \label{fig:msi}
\end{figure}

\paragraph{Coherence Protocol Examples.}
As mentioned above, there are many specific instances of cache coherence protocols.
Figure~\ref{fig:msi} illustrates the state transition diagram of MSI~\cite{censier.tc1978}, a popular coherence protocol.
MSI takes its name from the three possible states a cache block can be in:
(1)~\emph{modified}, which gives a core permissions to read from \emph{and} write to the cache block;
(2)~\emph{shared}, which gives a core permissions to read from the cache block; and
(3)~\emph{invalid}, which means that the cache block is not in the cache (and the core that the cache corresponds to therefore cannot read from or write to the block).
For the sake of simplicity, let's examine these states for a system with four CPU cores, where each core has its own private L1 cache, and the cores all share a single L2 cache.
When a cache block is in the shared state, multiple L1 caches can hold identical copies of the cache block.
Since none of the cores have permissions to write to the cache block, this ensures that any core reading the cache block has the most recent version of the data.
When a thread running on one of the cores wants to write to the cache block, it first sends a message to the other cores, informing them to downgrade to the invalid state (i.e., to invalidate their copy of the cache block, if they have one).
Once the core receives acknowledgments of the invalidations, it upgrades its own copy of the cache block to the modified state.\footnote{While not shown, practical cache coherence protocols implement additional \emph{transient states}, to indicate an in-progress upgrade/downgrade while waiting for other cores to complete their requested state changes.}
This now ensures that only this core has a copy of the cache block, and that the core can now safely perform its write.
If another core wants to read the now-modified cache block, it will send out a read request, which will force the writing core to downgrade its cache block from the modified state to the shared state, and to make the updates visible to the other cores.

\begin{figure}[!h]
    \centering
    \includegraphics[width=0.87\linewidth]{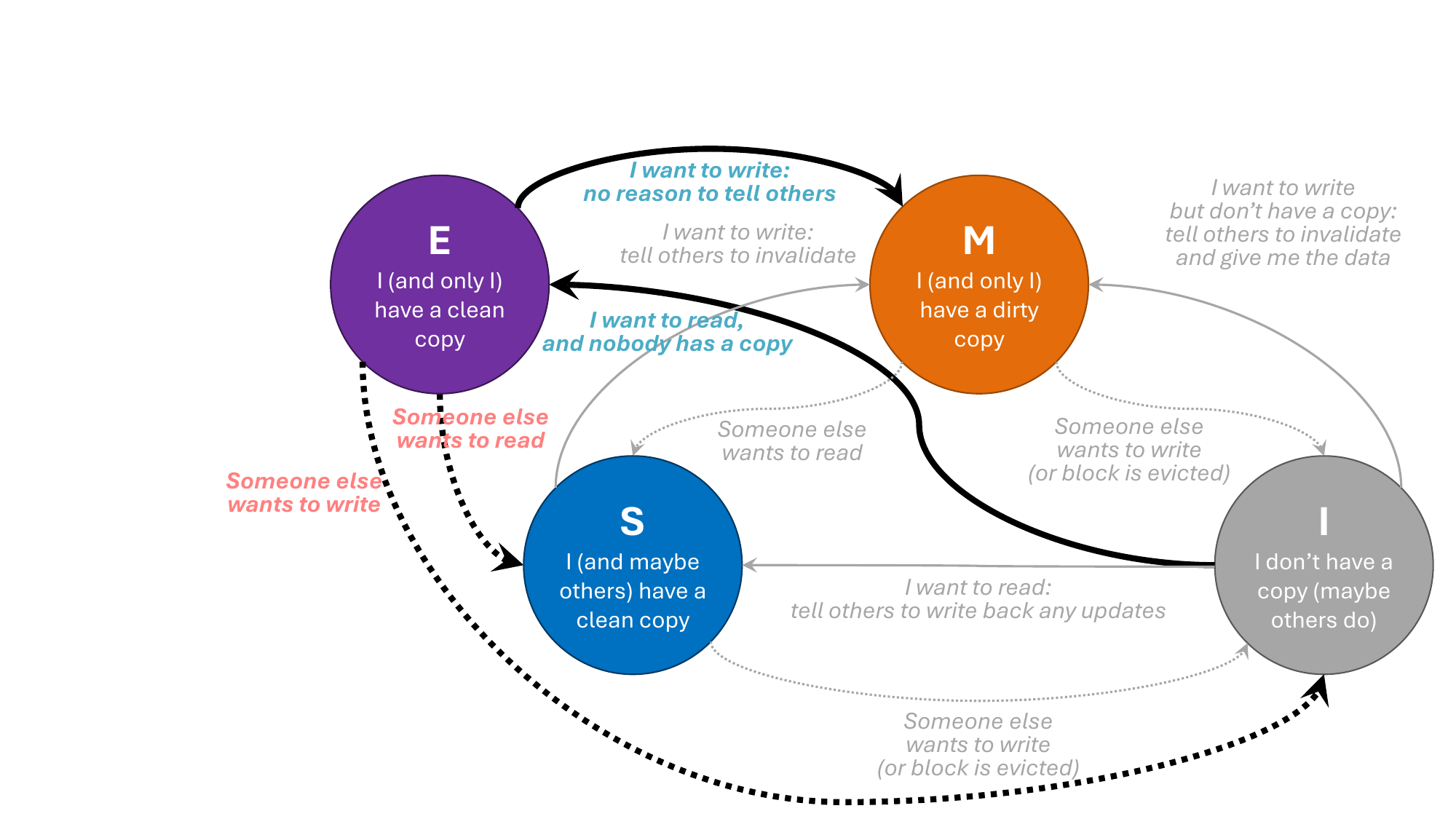}
    \caption{MESI protocol: solid lines are upgrades, dotted lines are downgrades. Transitions that are unchanged from MSI are grayed out.}
    \label{fig:mesi}
\end{figure}

One drawback of MSI is that even when only one core has a copy of the cache block in the shared state, it must wait for acknowledgments from all of the other cores before it can upgrade this cache block to the modified state.
This can potentially generate a large amount of unnecessary coherence traffic, given the lack of other copies in the private caches.
The MESI cache coherence protocol~\cite{papamarcos.isca1984} provides an optimization for this drawback, by introducing a fourth state called \emph{exclusive}, as shown in Figure~\ref{fig:mesi}.
If a core reads a cache block into its private cache, and no other private cache has a copy, the cache block will have an exclusive state, indicating that the core can read from the block, \emph{and} that no other private copy exists.
If the core subsequently wants to perform a write to the block, it can now silently upgrade the cache block (i.e., without sending any coherence messages to other cores) to the modified state.
To ensure correctness, if a second core wants to read the cache block, the first core's copy is downgraded from exclusive to shared, indicating that there may be more than one core currently holding a copy of the block.

\subsection{Memory Consistency Models}
\label{sec:mem:consistency}

A key consideration in coordinating memory updates across cores is the need to present a globally consistent view of these updates to all cores.
While cache coherence ensures that updates made by any one core to one unit of data are visible immediately to all cores, memory consistency~\cite{nagarajan.book2020} deals with how requests are interleaved across multiple units of data.
Unlike cache coherence, which is a purely \emph{microarchitectural} technique (i.e., transparent to the programmer) because it is a result of only microarchitectural design choices, memory consistency is an \emph{architectural} technique that must be exposed to the programmer, because data communication across threads is determined by the programmer in both implicit and explicit ways. 
There are many different memory consistency models, where each memory consistency model defines which possible memory interleavings can be observed by a core.
Regardless of the consistency model, each core must maintain an ordering of loads and stores that does not violate true (i.e., read after write) register dependencies, and because caches are part of the shared-memory state, cache coherence is implemented.
This section examines three popular types of consistency models.

\emph{Sequential consistency} (SC)~\cite{lamport.tc1979} ensures that \emph{every core} sees the same ordering (i.e., interleaving) of individual memory operations.
This is the equivalent of cores going one at a time when performing a load or store, and that load or store is considered part of the shared-memory state so that all cores see the effect (though not all cores need to see the effect immediately).
SC is often thought of as one of the most intuitive models, but given its need for a globally consistent interleaving, it can require a high overhead for implementation, and is rarely used for modern multicore CPUs.

\emph{Relaxed memory consistency models} allow for some differences in orderings observed by each core.
One example of a relaxed model is \emph{total store ordering} (TSO)~\cite{sparc.sparcv8.manual.1991}, which was first developed for the SPARC ISA and is used widely (albeit with some modifications) by the x86 ISA.
In TSO, a CPU core can observe its own write before other cores can, resulting in a slightly different interleaving observed by each core.
A key goal of TSO is to retain \emph{store buffers} in multiprocessors.
For an out-of-order core, a store buffer holds stores that have been committed, but have yet to be written to the caches.
In SC, the store buffer is not considered part of the shared-memory state, and memory speculation techniques that read data from the store buffer can require costly mechanisms to squash and replay out-of-order load that violate the global interleaving.
With TSO's relaxation, there is no need for such costly mechanisms, and the write buffer is considered part of the shared-memory state.

An example of an even more relaxed model is \emph{weak ordering}~\cite{dubois.isca1986}.
Weak ordering allows most memory operations to be reordered,
but uses programmer-invoked synchronization primitives to explicitly define reordering boundaries.
A popular synchronization primitive for weak ordering is a \emph{fence}, which is used to enforce orderings within a core (but does not explicitly synchronize across cores, unlike a barrier).
A fence has three guarantees:
(1)~all cores see the same exact ordering of fence primitives,
(2)~all load and store instructions that come before the fence in a thread must complete before the fence, and
(3)~no load or store instruction that comes after the fence can complete until after fence takes place.
What this means is that for the loads and stores in between two fences, any ordering of them can possibly occur.
The programmer can explicitly insert more fences into a thread to enforce a stricter ordering.
The Arm ISA is an example architecture that makes use of weak ordering.

Please refer to a detailed discussion~\cite{nagarajan.book2020} for more information about these and other memory consistency models.

%% file: sections/software.tex
\section{Optimizing Operating Systems for Multicore CPUs}
\label{sec:sw}

As mentioned in Section~\ref{sec:motivation}, multicore CPUs can make use of two types of parallelism: multiprocessing of multiple applications, and thread-level parallelism (TLP) within an application.
In order to best exploit these types of parallelism on multicore CPU hardware, operating systems and user applications have evolved in a number of key ways.
While these changes are not required to make use of multicore CPUs, they can allow developers to significantly improve the overall performance and efficiency of the system.
This section briefly touches on operating system changes and optimizations for multicore CPUs.

To facilitate our explanations, these software optimizations will be explained from the perspective of \emph{threads}.
As a simple broad definition, a thread is a sequence of CPU instructions from a program, and serves as the basic unit of scheduling for the operating system.
Generally speaking, when a program starts executing, that specific executing instance is known as a \emph{process}, and a process can consist of one or more threads of execution.

\paragraph{Scheduling Threads.}
\label{sec:sw:scheduling}
One of the many tasks that an operating system (OS) is responsible for is scheduling threads from all currently-running processes.
For a single-core CPU without multithreading, the CPU can execute only one thread at any given time.
As discussed in Section~\ref{sec:motivation:multiprocessing}, the OS maintains the illusion of concurrent thread execution by time-sharing the CPU.
At the end of a scheduling quantum, the OS uses a predefined \emph{scheduling policy} to select which of the active threads will execute on the CPU for the next quantum.
For the sake of simplicity, let us assume for now that the OS selected a new thread that is just starting its execution.
The selected thread is then assigned to the CPU's \emph{hardware context}, 
and all of the CPU's \emph{state} for that thread (e.g., program counter, registers, predictor history) are initialized to the starting state.
The thread then executes until the end of the quantum, unless the thread is preempted early (e.g., to execute an exception handler).

Assuming the typical conditions that no early preemption took place and that the thread has not yet finished executing, at the end of the quantum, the OS invokes the thread scheduler to select the thread for the next quantum.
If the selected thread is different from the currently-running thread, the OS
invokes a preemption of the running thread, which is known as a \emph{context switch}.
During a context switch, the OS copies the CPU state associated with the thread being preempted into memory, 
and loads in the CPU state for the thread being run next from memory.
From the perspective of the thread, the context switch gives the thread the illusion that it never stopped executing.
From the perspective of a user, PCs can often give them the illusion that all of their threads are running concurrently on a single-core CPU, by rapidly context switching between them, as the scheduling quantum is on the order of milliseconds, and is imperceptible to humans for the typical number of concurrently-running threads.

To extend thread scheduling for multicore CPUs, the individual cores are each exposed to the OS.
As an $n$-core CPU without multithreading has $n$~hardware contexts, the OS can schedule $n$~threads for every quantum.\footnote{When a CPU supports $m$-way multithreading, each way is typically exposed to the OS as its own hardware context, meaning that a $n$-core CPU with $m$-way multithreading exposes $n \times m$ hardware contexts.}
Conventionally, OS schedulers treated hardware contexts as identical to one another, but this introduces two challenges in modern multicore CPUs.

First, while a context switch preserves and restores CPU state for a thread, this state does not typically include the contents of the cache, because cache blocks are meant to be quickly-accessible copies of data that is available in other parts of the system.
However, it can be beneficial for a thread to be reassigned to a hardware context that it was previously scheduled on, as the thread can take advantage of data that it had previously cached.
Conventional scheduling approaches would disregard this, and assign the thread to \emph{any} available context.
\emph{Processor affinity}, sometimes referred to as \emph{CPU pinning}, overcomes this by allowing a user to assign a thread for execution on only the user-specified hardware contexts (e.g., the user assigns a thread to only one CPU core).
The OS scheduler obeys the processor affinity assignments, ensuring that the thread executes only on the selected contexts.

Second, as discussed further in Section~\ref{sec:modern:hetero}, modern multicore CPUs no longer have homogeneous cores.
As a result, the choice of hardware context to assign to a thread can have a significant impact on its performance and energy consumption (e.g., assigning a thread to a big core when it needs only a little core).
Early systems with two types of cores made the difference between cores transparent to the OS:
a single hardware context was associated with both a big core and a little core, and the hardware would use the CPU frequency setting chosen by the OS for the CPU (see \emph{Controlling CPU Core Frequency} below) to decide whether the thread should run on the big core or on the little core.
Modern systems can have more than two types of CPU cores, and often expose all of the CPU cores directly to the OS.
To manage these cores efficiently, OSes now include heterogeneity-aware schedulers, such as the Energy Aware Scheduling approach available in modern versions of the Linux kernel~\cite{linuxkernel.energyaware.manual}.

\paragraph{Controlling CPU Core Frequency.}
As introduced in Section~\ref{sec:hw:cores}, multicore CPUs expose the ability to perform per-core dynamic voltage and frequency scaling (DVFS).
In modern systems, the voltage and frequency setting for each core is chosen by the OS.
CPU manufacturers expose the DVFS capabilities as a series of frequency steps, where the OS chooses a target frequency based on certain properties, and the CPU uses the selected frequency step to control the voltage and frequency of the core.

While specific implementations vary, this section will focus on the Linux kernel implementation due to its readily-available documentation.
Linux includes a series of \emph{governors}~\cite{linuxkernel.governor.manual}, which are policies that the OS uses to select a frequency setting for a CPU.
Some governors, such as \emph{performance} or \emph{powersave}, constantly run a CPU at a fixed frequency (for the two examples, maximum or minimum, respectively).
Other governors, such as \emph{ondemand} and \emph{conservative}, use basic thread statistics such as load and idle time to select the frequency.
Newer governors, such as \emph{schedutil}, extract information about which threads were scheduled for the upcoming scheduling quantum to determine the frequency.
In addition to these built-in governors, developers can create custom governor policies for their systems.
A recent addition to the Linux kernel, Capacity Aware Scheduling~\cite{linuxkernel.capacityaware.manual}, combines core heterogeneity information with frequency settings chosen by the governor to guide hardware context assignment during thread scheduling.

\paragraph{Parallelizing Programs.}
\label{sec:sw:pgm}
On the application side, there are two ways to take advantage of the parallelism and efficiency offered by multicore CPUs.
The first method is to rewrite programs as \emph{multithreaded applications}.
Instead of writing a program as a fully-sequential series of functions, a programmer can identify opportunities to perform some parts of the application concurrently.
The programmer can use threading libraries to either 
(1)~explicitly spawn these parallel parts into independent threads, or
(2)~demarcate regions of the program that inform an advanced library to automatically generate threads.
Note that threads do not have to be fully independent, and that programmers can use \emph{synchronization primitives} to coordinate execution across threads (e.g., locks to protect critical section execution, barriers to synchronize task or sub-task completion), as well as \emph{shared memory} or \emph{message passing} to exchange information between threads.
While this chapter does not go into multithreaded programming in detail, there are other works~\cite{mattson.book2004,farooqi.bookchapter2022} that provide in-depth coverage of parallel programming techniques and frameworks.

The second method is often known as \emph{multiprogrammed execution}, where multiple independent processes execute concurrently on the CPU.
With multiprogrammed execution, it is possible to use all of the cores in a multicore CPU even if all of the applications are single-threaded, by allowing the processes to execute in parallel.
Most operating systems enable users to launch multiple processes concurrently, and when a process starts (initially with one thread), the OS adds the process' thread to the list of all active threads for the OS to schedule (see \emph{Scheduling Threads} above).
If a process is multithreaded, it will spawn additional threads over time, which the OS will also add to its list of active threads.
The OS scheduler will typically not distinguish between threads belonging to the same process and threads from other processes, and will schedule as many threads as there are hardware contexts.

%% file: sections/metrics.tex
\section{Evaluating Multicore CPUs}
\label{sec:metrics}

While conventional metrics such as speedup have been widely used in the architecture community for decades, several challenges make them difficult to directly apply to multicore CPU evaluations.
This section provides a summary of key challenges for performance measurement, and then discuss popular metrics that overcome these challenges.
It also briefly discusses metrics related to power and energy, given their emphasis throughout the lifetime of multicore CPUs.
\todo[mention simulators and perf counters?]

\paragraph{Multithreaded Application Performance.}
During the evaluation of multithreaded applications, an important challenge is factoring in the synchronization overhead.
Due to the non-deterministic nature of synchronization, multiple runs of the same program on the same machine, with identical data sets and an identical number of threads, can have different total runtimes, and can have different total instruction counts (e.g., a thread spinning as it waits to acquire a lock will execute additional instructions for each time it checks the lock variable).
As a result, it is important to 
(1)~compare equivalent amounts of work performed, as opposed to specific instruction counts (especially in architectural simulators that tend to execute only parts of a program); and
(2)~execute the configuration for each data point multiple times, and report a mean (ideally with error bars) instead of the results of a single run.
\todo[clean up first part?]

For a fixed amount of work (e.g., an entire application, a multithreaded kernel such as an entire parallel section of an application), the best measurement for comparison is the total wall clock execution time (i.e., the time from when the application/kernel starts, to the time the \emph{last} thread finishes).
While CPU cycle counts can act as a proxy for execution time, one must take care to use the global cycle count, and not a per-thread cycle count that might not track time during which a thread went to sleep (e.g., to wait for a lock).
To understand the benefits of parallelism when a program uses $N$ CPU cores, a common metric that is used is \emph{parallel speedup}, $S(N)$:
\begin{equation}
    S(N) = \frac{T_{s}}{T_{p}}
    \label{eq:parallelspeedup}
\end{equation}
where $T_p$ is the total execution time of the parallel version of the program for $N$~cores, and $T_s$ is the total execution time of the \emph{sequential} version of the program (and not the parallel version with one core).
The equation uses the sequential version of the program to ensure that parallel speedup captures the overheads of synchronization.
As a result, $S(1)$ can often be less than 1.
Note that unlike with single-thread applications, the IPC (\emph{instructions per cycle}) of a program should not be used as a substitute for execution time, as IPC values can be skewed by synchronization traffic and other non-deterministic behavior.

A related metric is \emph{parallel efficiency}, $E(N)$:
\begin{equation}
    E(N) = \frac{T_{s}}{T_{p} \times N}
    \label{eq:parallelefficiency}
\end{equation}
For parallel efficiency, a value of 1 indicates no overheads due to parallelization (i.e., the application is making full use of all cores), whereas values significantly lower than 1 indicate high overheads due to issues such as synchronization or serial execution.

\paragraph{Multiprogrammed Workload Performance.}
\emph{Multiprogrammed workloads} (i.e., a bundle of independent, concurrently executing applications) are unable to employ the performance metrics used by multithreaded applications because the different applications are working to complete separate tasks.
As a result, the applications in a workload can exhibit significant heterogeneity.
For example, a four-application workload may have two of the applications that are compute-intensive (and thus have high IPCs), while the other two applications are memory intensive (and thus have low IPCs).
If one were to use an aggregate metric of total execution time or the sum of IPCs (e.g., for comparing several potential system improvements to decide which one to implement), the metric may unfairly bias improvements to one class of applications (e.g., compute-intensive applications) over the other.
To avoid this bias when measuring overall system performance for a multiprogrammed workload, it is important to use metrics that attempt to \emph{normalize} the relative benefits to each application, though there are several ways to perform this normalization~\cite{eyerman.ieeemicro2008}.

Related to this is the issue of \emph{fairness}.
As the applications in a multiprogrammed workload are independent of each other, one application's resource usage can generate interference that slows down the other applications that are sharing the multicore CPU, compared to if those applications ran alone on the CPU.
For any one application in the workload, its \emph{slowdown} is defined in terms of its performance when running \emph{alone} on the CPU compared to its performance when the CPU is \emph{shared} with other applications~\cite{kim.pact2004, mutlu.micro2006}:\footnote{To simplify the discussion of multiprogrammed workload metrics, this section assumes that each application is single-threaded, which allows us to use IPC in the equations (as there is no synchronization overhead). For multithreaded applications in a multiprogrammed workload, IPC should be replaced by total execution time for that application.}
\begin{equation}
    slowdown = \frac{IPC_{alone}}{IPC_{shared}}
    \label{eq:slowdown}
\end{equation}
By using a ratio of IPCs, the slowdown metric normalizes away the inherent compute-intensive or memory-intensive behavior of each application, and reports what fraction of the application's performance was lost due to interference.
A system is \emph{unfair} if it slows down some applications significantly more than others.

There are two metrics that can quantify (un)fairness.
The first is \emph{maximum slowdown}, which simply says that the worst slowdown experienced by any one application is an indication of unfairness due to interference, and that a smaller maximum slowdown is more equitable~\cite{das.micro2009}.
The second defines fairness as the ratio between the best slowdown and worst slowdown experienced by applications in the workload~\cite{gabor.micro2006, eyerman.ieeemicro2008}:
\begin{equation}
    fairness = \min_{i, j} \frac{slowdown_i}{slowdown_j} = \frac{\min_{i} slowdown_i}{\max_{j} slowdown_j}
\end{equation}
where $i$ and $j$ are members of the set of all applications in the workload.
Informally, a fairness of 1 indicates that all applications are experiencing an equal slowdown, while at the other extreme, a fairness of 0 indicates that at least one application is experiencing starvation.
Note that unfairness in this case is defined as the inverse of fairness.

Aggregate performance metrics for multiprogrammed workloads incorporate some notion of both overall system throughput and fairness, in an attempt to remove the bias mentioned above that can arise from using absolute IPCs.
A popular metric is weighted speedup ($WS$)~\cite{snavely.asplos2000, eyerman.ieeemicro2008}, which sums up the normalized speedups (i.e., the inverse of slowdown) of each application $i$ in the workload to represent system throughput:
\begin{equation}
    WS = \sum_{i} \frac{IPC_{shared}}{IPC_{alone}} = \sum_{i} \frac{1}{slowdown_i}
    \label{eq:weightedspeedup}
\end{equation}
A larger value of $WS$ is better, as it indicates higher system throughput (i.e., lower aggregate impacts of interference).
Note that for an $n$-application workload, $WS$ typically ranges between 0 and $n$, indicating that the metric is dependent on the number of applications (and often, by proxy, the number of cores), as this represents a throughput  for a specific system.
As a result, when comparing two systems and reporting \emph{improvements} due to a system modification, one should typically report a ratio of weighted speedups (occasionally referred to as \emph{WS improvement}, although this is sometimes confusingly reported as just WS in several papers):
\begin{equation}
    WS\ \text{improvement} = \frac{WS_{after\_modification}}{WS_{before\_modification}}
    \label{eq:wsimprovement}
\end{equation}

There is some debate about whether $WS$ effectively captures unfairness, which has led several researchers to use the \emph{harmonic mean} ($HM$) of speedups~\cite{luo.ispass2001, eyerman.ieeemicro2008} (sometimes referred to as \emph{harmonic speedup}) in place of or in addition to $WS$:
\begin{equation}
    HM = \frac{n}{\sum_{i} \frac{IPC_{alone}}{IPC_{shared}}} = \frac{n}{\sum_{i} slowdown_i}
    \label{eq:harmonicmean}
\end{equation}
$HM$ represents the average slowdown in a user response (i.e., the turnaround time for an output produced by the application) for each application in the workload due to interference~\cite{eyerman.ieeemicro2008}.
Like with $WS$, higher is better, and to compare two systems and report an improvement, one should calculate the ratio of $HM$ values for the two systems.

\paragraph{Power and Energy.}
While power and energy are related, they represent different limiting factors experienced by modern computers.
Power ($P$) represents a \emph{rate} of work being completed, and can be calculated as a function of current ($I$) and voltage ($V$):
\begin{equation}
    P = I \times V
    \label{eq:power}
\end{equation}
At a high level, power in a CPU can be broken down into a \emph{dynamic} component (e.g., the power consumed due to the active switching of transistors to perform work, short-circuit power consumed during gate switching when transistors temporarily connect the high voltage rail to ground due to transistor timing variation) and a \emph{static} component (e.g., the leakage of power due to imperfections in the switching behavior of a transistor).
Section~\ref{sec:hw:cores} briefly discusses components that impact dynamic power.
Note that in the past, dynamic power was orders of magnitude larger than static power, so static power was thought of as a  trivial factor in total power consumption. Today, because decades of Dennard scaling translated to significant dynamic power reductions, static power makes up a non-trivial fraction of total CPU power.
Power consumption has a direct correlation with thermal dissipation, and is used as a proxy to quantify the heat generated by the CPU.
The \emph{areal power density}, which divides the power by the surface area of the CPU die, is used to determine how aggressive thermal cooling solutions (e.g., heatsinks, liquid cooling, fans) need to be to remove dissipated heat from the CPU and keep the die within safe thermal operating limits.
While areal power density is used at design time to provision heat dissipation capacity,
the CPU makes use of \emph{temperature} readings from multiple sensors embedded at various locations in both the CPU chip and the motherboard to dynamically control heat dissipation management, including cooling intensity (e.g., fan speed) and CPU power throttling.

While power consumption was the key concern during the early years of multicore CPUs, energy emerged as a first-order concern during the 2010s.
Energy ($E$) is the total electrical cost of performing a given amount of work, and can be calculated as a function of power:
\begin{equation}
    E = P \times t
    \label{eq:energy}
\end{equation}
where $t$ is the time required to complete the defined amount of work.
Challenges associated with two extreme ends of computing platforms have resulted in the growing emphasis on energy consumption, in addition to power consumption.
First, portable computers such as laptops and smartphones are battery-constrained, as their available uptime depends on the total energy capacity of the computer's battery and the amount of energy that the system (including the CPU) consumes for running applications.
Second, the large number of servers in data centers and cloud computing environments can result in exorbitant financial and environmental costs to provide enough energy to perform user services.
In both cases, reducing the total energy used for a given application can result in more availability at a lower overall cost.
A related metric of interest is \emph{energy efficiency}, which summarizes the energy used for a single operation (e.g., an instruction, a microkernel), though one downside of energy efficiency is the difficulty of defining an operation in an equal way across platforms (e.g., two CPUs with different ISAs may not have equivalent instructions).
Several modern multicore CPUs contain ISA-compatible cores of heterogeneous size and capability (see Section~\ref{sec:modern:hetero}), and dynamic energy and/or energy efficiency metrics based on the characteristics of a thread are used to select which of the heterogeneous cores will execute the thread.

%% file: sections/modern.tex
\section{The Evolution of Multicore CPUs}
\label{sec:modern}

In the years that have elapsed since the introduction of multicore CPUs, there have been a number of innovations to the general microarchitecture described in Section~\ref{sec:hw}.
While some of these innovations have been limited to specific manufacturers or models, others have become commonplace across modern CPUs.
This section highlights three of the most significant shifts in multicore CPU design, and leaves the exploration of other innovations as an exercise for the reader.
These three shifts are finding widespread acceptance in contemporary multicore CPUs:
(1)~the integration of specialized components on-chip alongside the general-purpose CPU cores into what are known as \emph{systems-on-chip} (SoCs);
(2)~the diversification of the constituent cores in a multicore CPU; and
(3)~the advent of composable \emph{chiplets} that can allow for the easy integration of many smaller silicon dies in a single chip.

\subsection{Systems-on-Chip}
\label{sec:modern:soc}

Just as the limits of areal power density and thermal dissipation were a key motivator for the rise of multicore CPUs, a new pressure point that came to prominence a few years later ushered in the next key change.
The emergence of the smartphone drove a need to reduce total energy consumption, given the limited battery capacities that were available in a portable form factor.
To maximize energy efficiency, smartphones made use of \emph{systems-on-chip} (SoCs).
A system-on-chip tightly integrates many different components, which conventionally would have been implemented using multiple chips for desktop and server computers, into a single chip.
Early SoC examples date back to the mid-1970s, such as the Intel 5810~\cite{intel.5810a.datasheet.1976} introduced in 1974, and were designed to minimize battery consumption in then-new electronic wristwatches.
Over time, platform-specific SoCs became relatively commonplace in the embedded systems community.
Early smartphones such as the original Apple iPhone, from 2007, made use of Samsung SoCs that contained a single Arm CPU core, a graphics processing unit (GPU), and caches integrated onto a single chip~\cite{mannion.eetimes2007}.

\begin{figure}[h]
    \centering
    \includegraphics[width=0.52\linewidth]{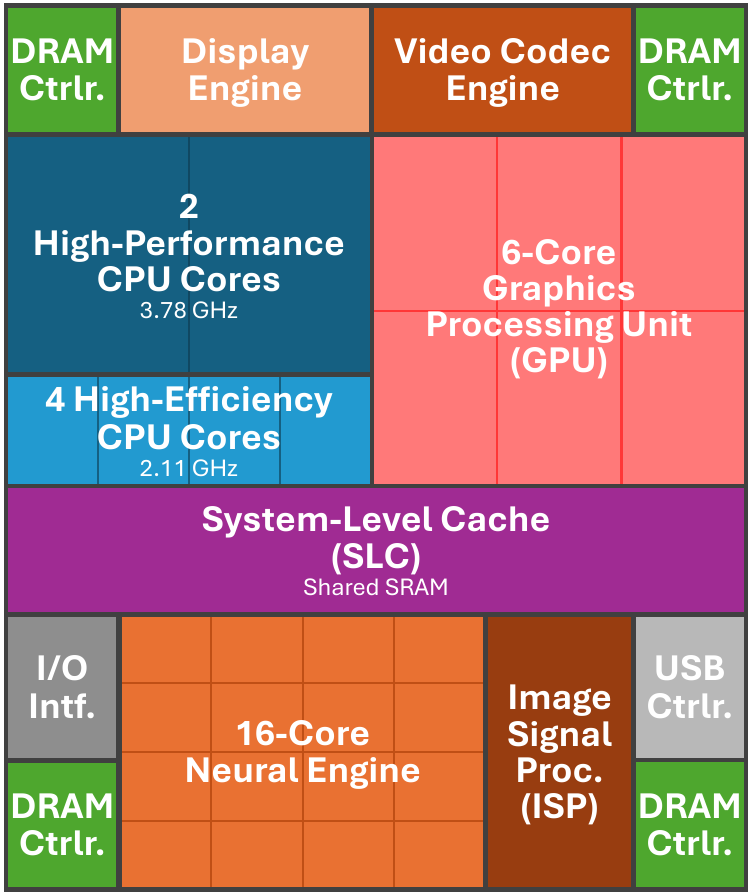}
    \caption{Selected components in the Apple A17 Pro system-on-chip. Note that components may not be to scale with each other.}
    \label{fig:a17pro}
\end{figure}

As the functionality and ubiquity of the smartphone expanded, their underlying SoCs incorporated significantly more components, including more CPU cores.
At a high level, the goal of these additional components is to introduce \emph{specialization} for commonly-performed operations, in order to significantly improve the efficiency of these operations compared to executing them on a general-purpose CPU core.
Figure~\ref{fig:a17pro} shows several key components of the Apple A17 Pro SoC~\cite{apple.a17pro.event2023}, which started production in 2023 for use in the iPhone 15 Pro series of smartphones.
As the figure shows, the SoC contains six CPU cores (which are heterogeneous; see Section~\ref{sec:modern:hetero}), a GPU, multiple fixed-function accelerators (a neural engine for machine learning inference, an image signal processor for photo processing, a video codec engine for video recording and streaming, a display engine for screen image generation), I/O interfaces (including a dedicated USB controller), a system-level cache (an LLC that is available to the CPU, GPU, and all accelerators), and four LPDDR5X memory controllers.

The process of identifying which components to include in the SoC (beyond the basic CPU and GPU) makes use of \emph{profiling tools}.
These tools can monitor the performance (and often the energy) of an existing chip as one or more applications execute.
To do this, modern profiling tools make use of hardware \emph{performance counters}, which are registers built into the chip logic to track various events taking place during execution.
For SoC design, profiling can identify the applications or application kernels that are bottlenecked by the existing chip, which become candidates for acceleration using dedicated SoC components.
Of these candidates, a subset of them are chosen for dedicated acceleration based on a combination of factors, including the frequency of application/kernel usage, the area required for a fixed-function accelerator, available chip area, power and energy budgets, and the availability of existing accelerator designs (including the availability of third-party designs known as \emph{IP cores}, where IP stands for intellectual property).
Note that even among state-of-the-art SoCs that target the same platform, the exact components can vary.
As one example, Qualcomm's Snapdragon~8 Gen~3 SoC~\cite{qualcomm.snapdragon8gen3.brief2024} for smartphones integrates a 5G cell modem, WiFi and Bluetooth transceivers, and security accelerators in the chip, and makes use of a different combination of heterogeneous cores than the A17 Pro.

\subsection{Heterogeneous CPU Cores}
\label{sec:modern:hetero}

As discussed in Section~\ref{sec:modern:soc}, a key motivator for introducing fixed-function accelerators in SoCs is to improve energy efficiency for commonly-performed operations.
Even with these accelerators, there remains a need to execute code using a general-purpose CPU core (e.g., less common operations, OS system calls, irregular workloads that are more difficult to accelerate).
Using the smartphone as a motivating example, the CPU cores in a smartphone SoC have seen significantly increasing complexity in their designs (e.g., more sophisticated predictors, deeper speculation, larger superscalar widths~\cite{grayson.isca2020}), to meet the increasing demands of smartphone applications that have become more sophisticated over the years.
While larger CPU cores can enable efficient execution of modern compute-heavy applications compared to smaller cores, the increased core complexity (along with associated increases in leakage and clock power) become more \emph{inefficient} for more memory-bound applications that cannot make effective use of these additional resources.

A seminal study in 2003 illustrated how having a \emph{heterogeneous} combination of large cores \emph{and} small cores in a single multicore CPU is significantly more energy efficient than the use of homogeneous (i.e., identical) cores~\cite{kumar.micro2003}.
If all of the cores maintain the same instruction set architecture (ISA), an application can migrate from a large core to a smaller core when it enters a more memory-bound phase, and can migrate back to the larger core when it enters a more compute-bound phase.
Arm introduced their first CPU cores designed for a heterogeneous multicore CPU in 2011, under the name big.LITTLE~\cite{greenhalgh.whitepaper2011}.
The first big.LITTLE pairing combined large Cortex-A15 cores with small Cortex-A7 cores, where the Cortex-A15 cores could achieve up to 3.0$\times$ the performance of the Cortex-A7 cores, but the Cortex-A7 could achieve up to 3.8$\times$ the energy efficiency, for selected benchmarks.
For what Arm defined as low- to mid-range workloads, the Cortex-A7 was expected to execute these workloads with a dramatic reduction in power consumption, with no expected loss in performance compared to the Cortex-A15.
Initially, core heterogeneity was hidden from the system software, with the CPU hardware transparently migrating applications between a big core and a paired little core.
This meant that for any big--little pair, one of the cores was idle.
Today, core heterogeneity is exposed to the OS scheduler, as discussed in Section~\ref{sec:sw:scheduling}.

Since the commercial introduction of heterogeneous CPU cores, they have become commonplace in modern multicore CPUs.
The example A17 Pro CPU from Apple, shown in Figure~\ref{fig:a17pro}, contains two types of CPU cores~\cite{apple.a17pro.event2023}:
(1)~a pair of large high-performance cores, and
(2)~four smaller high-efficiency cores.
Qualcomm's Snapdragon~8 Gen~3 CPU contains three distinct types of CPU cores~\cite{qualcomm.snapdragon8gen3.brief2024}:
(1)~a single large prime core that runs at a frequency of up to \SI{3.4}{\giga\hertz};
(2)~five performance cores, large but notably smaller than the prime core, that run at a frequency of up to \SI{3.2}{\giga\hertz}; and
(3)~two small efficiency cores that run at a frequency of up to \SI{2.3}{\giga\hertz}.
While the usage of heterogeneous CPU cores first became popular for mobile SoCs, it can now be seen in a wide range of modern multicore CPUs.
For example, Intel's 12th generation of Core CPUs introduced heterogeneous CPU cores (named P-cores and E-cores) for desktop computers~\cite{rotem.hc2021}.

\subsection{Chiplet-Based Multicore Design}
\label{sec:modern:chiplets}

Beyond the cores themselves, the success of SoCs have demonstrated the benefits of maximizing energy efficiency through directed specialization.
However, there remains a tension between high degrees of specialization and \emph{non-recurring engineering} (NRE) costs, such as those involved with design, layout, and verification.
As a simple motivating example, let's revisit the tiled multicore design from Section~\ref{sec:hw:cores}.
While tiling helps reduce NRE costs, the tile is a fixed design: for one core, there is a fixed amount of L1 instruction and data caches, L2 cache, and potentially LLC slice.
If a manufacturer wants to adapt this tile for a platform whose workloads do not exhibit significant locality, they may want to significantly reduce the cache sizes, but doing so requires a new tile to be designed and verified.
Moreover, the die that is etched for a multicore CPU will have a fixed number of tiles laid out, again restricting the flexibility of the CPU and requiring non-trivial NRE costs if a die with a different core/tile count is needed.

The advent of \emph{chiplets} provides a new way to compose a multicore CPU in a more modular fashion, avoiding some of these NRE costs.
A chiplet is a small die that contains a subset of the functionality that would be contained in a standard die.
Instead of laying out a multicore CPU design using a monolithic die, designers can design smaller chiplets with individual components, such as cores or caches.
For example, in place of a single die containing eight cores and their associated caches, a chip for a multicore CPU could be composed using eight core chiplets, eight L1 cache chiplets, eight L2 cache chiplets, and 16 LLC slice chiplets.
If the manufacturer now wants a multicore CPU with fewer cores and larger caches, they can reuse the chiplets to compose a chip with two core chiplets, 16 of the L1 and L2 cache chiplets each, and 64 LLC slice chiplets.
While chiplets are one example of a broader concept called \emph{multi-chip modules} (MCMs), which has been around for decades, it was conventionally difficult to have more than a handful of dies in an MCM due to packaging costs and alignment issues.
Recent advances in \emph{interposer} design, where an interposer provides a substrate with many short-distance wires to connect dies together, have reduced manufacturing costs, complexity, and faults for assembling many dies in one MCM.

Chiplets offer four advantages for manufacturing.
First, as already discussed, chiplets allow for modular components that can be reused and resized after the dies have been fabricated, at low cost.
Second, chiplets can allow for dies fabricated using different manufacturing process technologies to be connected together into a single package (this is known as \emph{heterogeneous integration}).
If, for example, the cache does not need to be manufactured using the state-of-the-art manufacturing process, manufacturers can reduce costs by fabbing the chip using an older, cheaper process.
Third, overall yield increases, because a silicon fault is now isolated to a much smaller chiplet, which can be replaced at much lower cost than disposing of an entire die.
Fourth, with the breakdown of Dennard scaling~\cite{dennard.iedm1972, dennard.jssc1974}, dies are now growing in size to continue scaling up the total transistor count, but these sizes are approaching the reticle limits (i.e., the largest possible chip that can be etched) of our lithography equipment.
Chiplets can overcome these limits by allowing for multiple larger chiplet dies to be composed into a package, where the total area of the chiplets is significantly larger than what any one die could be.

\begin{figure}[h]%
    \centering%
    \includegraphics[width=0.32\linewidth]{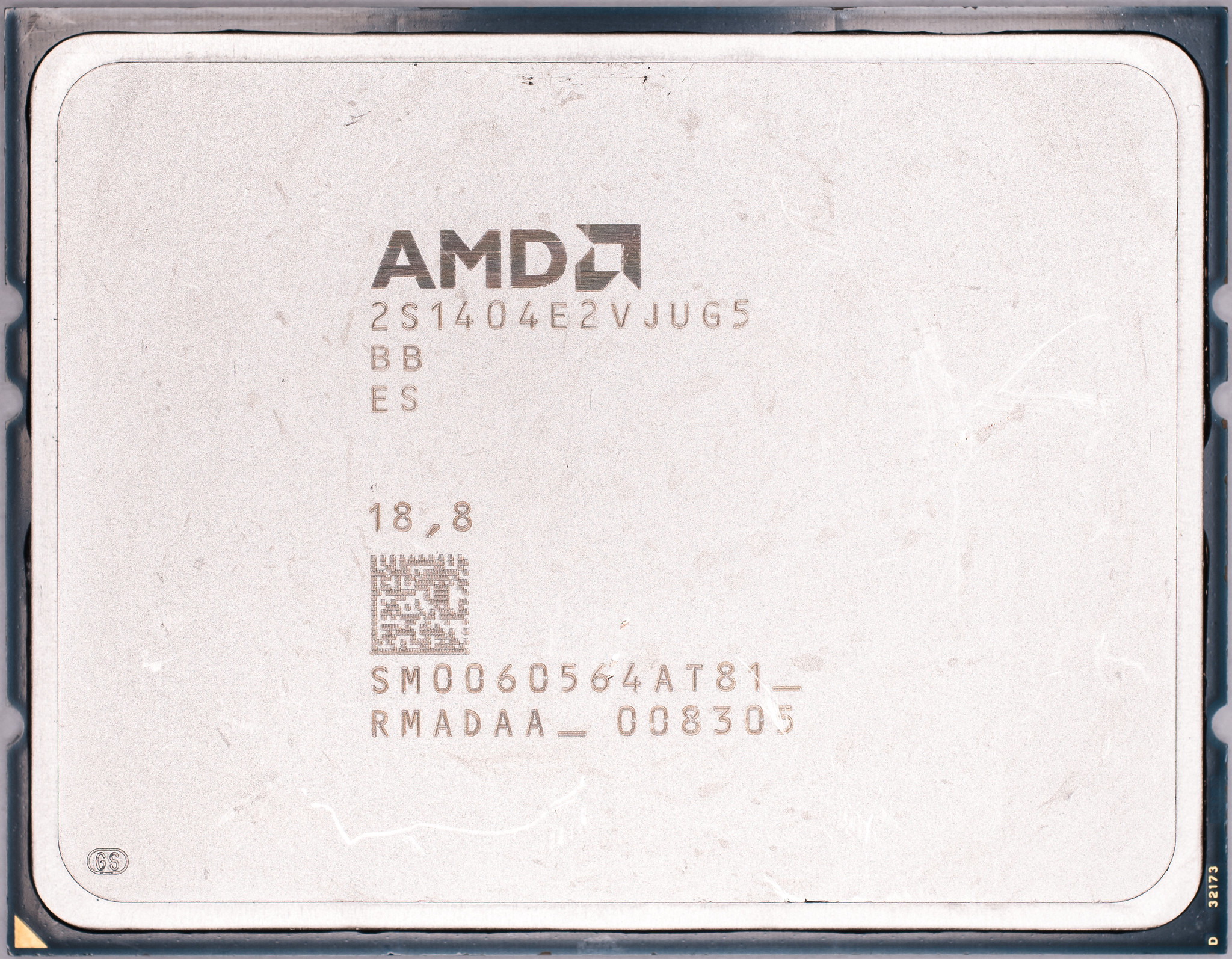}\hfill%
    \includegraphics[width=0.32\linewidth]{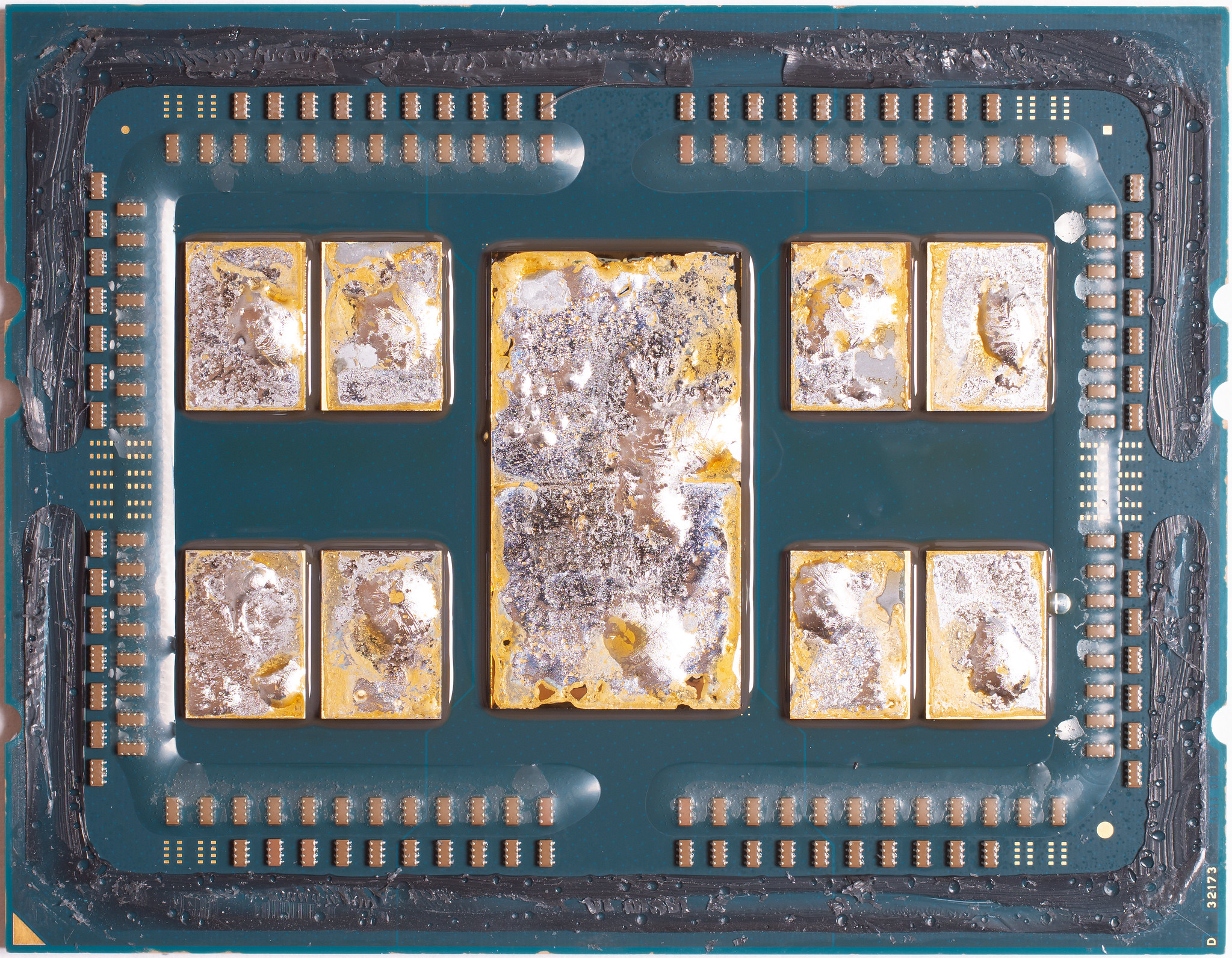}\hfill%
    \includegraphics[width=0.32\linewidth]{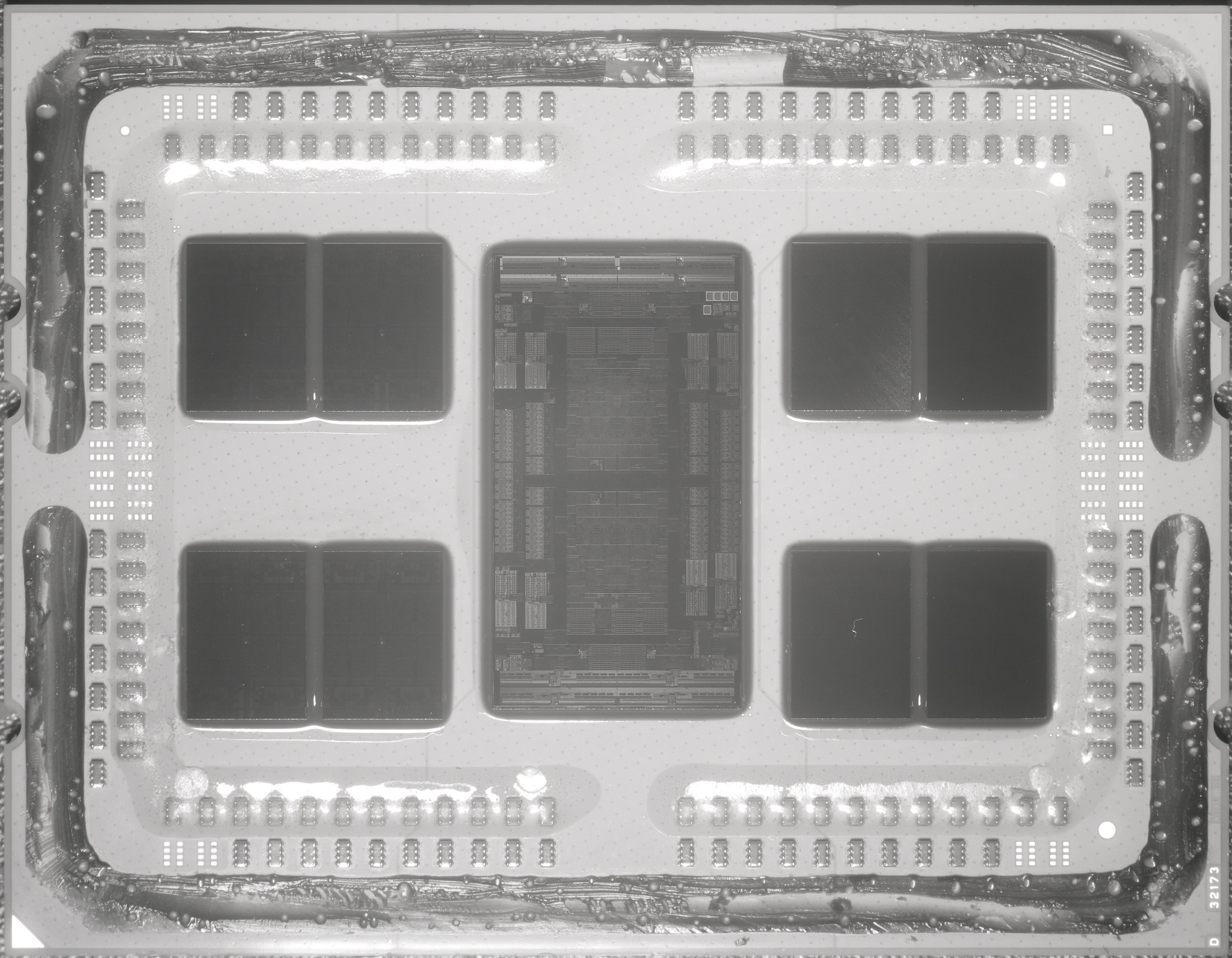}%
    \caption{Lidded, delidded, and infrared views of the AMD EPYC 7702 CPU and its chiplets~\cite{fritz.epyc7702.dieshots}.}%
    \label{fig:epyc}%
    \vspace{-10pt}%
\end{figure}

Several manufacturers have started incorporating chiplet-based design for multicore CPUs.
AMD has been responsible for significant innovation in the area of chiplets and interposer design, and has been manufacturing chiplet-based multicore CPUs starting with the first-generation EPYC CPUs in 2017~\cite{naffziger.isca2021}.
Figure~\ref{fig:epyc} shows die shots of the AMD EPYC 7702 CPU, released in 2019, which consists of nine chiplets:
eight core complex dies (CCDs), fabricated in a \SI{7}{\nano\meter} process, with eight cores (and their private caches and LLC slices) per CCD; and a single I/O die in the center, fabricated in a \SI{14}{\nano\meter} process, with memory and I/O controllers.
Apple and Intel have also announced the incorporation of chiplet-based design into their latest multicore CPUs~\cite{smith.spectrum2022, rodgers.ft2024}.

%% file: sections/conclusion.tex
\section{Conclusion}
\label{sec:conclusion}

The rise of multicore CPUs highlighted a key shift in the computer architecture community, as concerns about thermal dissipation and limitations of ILP hastened a collective change in mindset about the importance of power and the potential of parallel processing.
Innovations in multicore CPU design have led to reduced design and verification effort, support for specialized hardware accelerators, and increased composability of modular components.
Today, multicore CPUs can be found across most modern computers, ranging from embedded platforms and smartphones, through personal laptops and desktops, to large-scale distributed computing environments.
Combined with the emergence of simplified parallel programming frameworks, multicore CPUs have led to commonplace exploitation of thread-level parallelism, delivering significant performance improvements while maintaining reasonable power and energy budgets.
Multicore CPUs are expected to continue evolving over the next several decades, as recent trends towards CPU specialization (particularly in the modern CPU landscape with systems-on-chip and chiplet-based fabrication) open up new opportunities for maximizing the efficiency and performance of the next generation of computing platforms.